\documentclass[fleqn,usenatbib]{mnras}

\usepackage{newtxtext,newtxmath}

\usepackage[T1]{fontenc}

\DeclareRobustCommand{\VAN}[3]{#2}
\let\VANthebibliography\thebibliography
\def\thebibliography{\DeclareRobustCommand{\VAN}[3]{##3}\VANthebibliography}

\usepackage{graphicx}
\usepackage{amsmath}

\usepackage{amssymb}
\usepackage{mathtools}
\usepackage{gensymb}
\usepackage{textcomp, gensymb}
\usepackage{hyperref}
\usepackage{tablefootnote}
\usepackage{threeparttable}
\usepackage{xcolor}

\newcommand{\kms}{\mbox{$\mathrm{km\,s}^{-1}$}}

\newcommand{\Teff}{\mbox{$T_\mathrm{eff}$}}
\newcommand{\Msun}{\mbox{$\mathrm{M}_{\sun}$}}
\newcommand{\Rsun}{\mbox{$\mathrm{R}_{\sun}$}}

\newcommand{\Msunyr}{M$_{\odot}$\,yr$^{-1}$}

\newcommand{\Porb}{\mbox{$P_\mathrm{orb}$}}

\newcommand {\SDSSV}{SDSS\nobreakdash-V}

\newcommand{\obj}{SDSS\,J0852+7832}
\newcommand{\VAnd}{V479\,And}
\newcommand{\VSgr}{V1082\,Sgr}

\newcommand{\gaia}{{\it Gaia}}

\newcommand{\swift}{{\it Swift}}

\usepackage{orcidlink}

\title[Evolution of the donor star in SDSS\,J085210.48+783246.6]{Evolution of a long-period Cataclysmic Variable from the viewpoint of the donor star: the case of SDSS\,J085210.48+783246.6}

\author[Tovmassian et al.]{
G. Tovmassian \orcidlink{0000-0001-6208-9109}$^{1,2}$,\thanks{E-mail: gag@astro.unam.mx}
D. Belloni \orcidlink{0000-0003-1535-0866}$^{3},$
I. Mora Zamora \orcidlink{0000-0001-8600-4798}$^{4}$,
B.T. G\"ansicke \orcidlink{0000-0002-2761-3005}$^{5}$,
S. Zharikov \orcidlink{0000-0003-2526-2683}$^{1}$,
\newauthor J. Echevarria \orcidlink{0000-0002-2653-1120}$^{4}$,
M.R. Schreiber \orcidlink{0000-0003-3903-8009}$^{6}$,
P. D’Avanzo \orcidlink{0000-0001-7164-1508}$^{2}$,
P. Ochner \orcidlink{0000-0001-5578-8614}$^{7,8}$, 
R. Ashley\orcidlink{0000-0002-1210-4144}$^{5}$ and
K. Inight \orcidlink{0000-0002-2200-2416}$^{5}$
\\
$^{1}$Universidad Nacional Aut\'onoma de M\'exico. Instituto de Astronom\'ia. A.P. 106, 22800. Ensenada, B.C. , M\'exico\\
$^{2}$INAF – Osservatorio Astronomico di Brera, Via E. Bianchi 46, 23807 Merate (LC), Italy\\
$^{3}$International Centre of Supernovae (ICESUN), Yunnan Key Laboratory of Supernova Research, Yunnan Observatories, Chinese Academy of Sciences (CAS), Kunming 650216, China\\
$^{4}$Universidad Nacional Aut\'onoma de M\'exico. Instituto de Astronom\'ia. A.P. 70-264, 04510. Ciudad de M\'exico, M\'exico\\
$^{5}$Department of Physics, University of Warwick, Coventry, CV4 7AL, UK \\
$^{6}$Departamento de F{\'i}sica, Universidad T{\'e}cnica Federico Santa Mar{\'i}a, Av. Espa{\~n}a 1680, Valpara{\'i}so, Chile\\
$^{7}$Università degli Studi di Padova, Dipartimento di Fisica e Astronomia, Vicolo dell’Osservatorio 3, 35122 Padova, Italy\\
$^{8}$INAF -- Osservatorio Astronomico di Padova, Vicolo dell Osservatorio 5, I-35122 Padova, Italy\\
}

\date{Accepted XXX. Received YYY; in original form ZZZ}

\pubyear{\the\year{}}

\begin{document}
\label{firstpage}
\pagerange{\pageref{firstpage}--\pageref{lastpage}}
\maketitle

\begin{abstract}
Cataclysmic variables were long considered to be close binaries consisting of a white dwarf and a Roche-lobe-filling, near-zero-age main-sequence (ZAMS) red or brown dwarf. Recent massive surveys have uncovered an increasing number of binaries with similar spectral characteristics but harboring secondary stars that have undergone nuclear evolution and partial envelope stripping, many with orbital periods far exceeding the normal upper limits for ordinary CVs. We present a detailed study of a newly discovered CV with a 17.109\,h period and determine its basic stellar parameters. We also discuss the evolutionary paths leading to the formation of these extremely long-period cataclysmic variables. We consider the implications of the new evolutionary hypothesis on their further evolution into double-degenerate binaries.
\end{abstract}

\begin{keywords}
{cataclysmic variables --- binaries: close --- stars: evolution --- accretion, accretion discs --- white dwarfs --- stars: individual: SDSS\,J085210.48+783246.6}
\end{keywords}

\section{Introduction}\label{sec:intro}

In cataclysmic variables (CVs), close binary systems in which a white dwarf accretes from a main-sequence or slightly evolved donor star, the evolution proceeds from longer towards shorter orbital periods, driven mainly by magnetically guided winds carrying away angular momentum in a mechanism known as magnetic wind braking at periods above 3\,h and by gravitational waves at shorter periods \citep[e.g.][]{Knigge2011,Belloni+Schreiber_2023}.
For decades, the evolution of CVs has been modelled under the assumption of magnetic braking prescriptions derived from the spin-down of slowly rotating sun-like stars \citep{Skumanich_1972}. The resulting prescription predicts a strong dependence of angular momentum loss on the spin period \citep{Rappaport_1983}, and the corresponding evolutionary models are still considered the standard scenario for CV evolution.

However, there is growing evidence that this standard prescription
for magnetic braking is significantly flawed. Observational and
theoretical studies of fast-rotating low-mass stars have established
that the magnetic braking torque saturates for spin periods shorter
than a certain threshold; that is, the dependence of angular momentum
loss on rotation rate becomes much shallower than predicted by
\citet{Rappaport_1983} \citep[e.g.][]{Barnes+Sofia_1996,
Reiners+Mohanty_2012, Matt_2015}. \citet{El-Badry_2021} showed
that this saturation is also relevant in close binaries with evolved
donors, and that the standard prescription consequently
overestimates angular momentum loss in such systems.
Revised prescriptions that incorporate saturation have been successfully applied to CVs \citep{Barraza-Jorquera_2025} and their progenitors \citep{Belloni_2024}.

%Perhaps most importantly, \citet{Belloni+Schreiber_2023} showed that, assuming more %efficient angular-momentum loss via magnetic braking in slightly evolved donors, %long-standing problems in the formation of ultra-compact binaries from CVs can be %solved. {\bf Similar scenarios with even stronger braking are propagated by %\citet[]%[]{sarkaretal23-2, sarkaretal23-1}.}
%Assuming strong magnetic braking according to \citet{Van+Ivanova_2019}, AM\,CVn %systems, {\bf i.e. short orbital period ($\lesssim 60$\,min) binaries in which a %white dwarf accretes from a helium-dominated donor star, can form from CVs with %evolved donors.
%{\bf AM\,CVn systems, which are short-period ($\lesssim 60$\,min) binaries 
%in which a white dwarf accretes from a helium-dominated donor star, 
%can form from CVs with evolved donors assuming strong magnetic braking 
%according to \citet{Van+Ivanova_2019}. The same prescription is also 
%required to explain the formation of the previously termed paradoxical 
%double white dwarf binaries \citep{Aros-Bunster_2025, Belloni_2025}. These} might %be just a natural descendant of
%mass-transferring and recently detached CVs with evolved secondaries \citep{El-%Badry_2021}.

 Perhaps most importantly, \citet{Belloni+Schreiber_2023} showed 
that more efficient angular-momentum loss via magnetic braking in 
slightly evolved donors can solve long-standing problems in the 
formation of ultra-compact binaries from CVs. Under this picture, 
CVs with evolved donors and unusually long periods are the natural 
progenitors of extremely low-mass white dwarfs with more massive 
companions \citep{El-Badry_2021, Aros-Bunster_2025, Belloni_2025}. 
These in turn may evolve into AM\,CVn systems, short-period 
($\lesssim 60$\,min) binaries in which a white dwarf accretes from 
a helium-dominated donor \citep{Belloni+Schreiber_2023}. The 
prescription of \citet{Van+Ivanova_2019}, and the even stronger 
braking scenarios explored by \citet{sarkaretal23-2, sarkaretal23-1}, 
provide the angular-momentum loss rates required to connect these 
systems within a single evolutionary sequence. The previously termed 
paradoxical double white dwarf binaries \citep{Aros-Bunster_2025, 
Belloni_2025} fit naturally into this sequence as descendants of 
mass-transferring and recently detached CVs with evolved secondaries 
\citep{El-Badry_2021}.

%In other words, assuming strong magnetic braking allows us to model all the above-%mentioned systems within a single evolutionary sequence. CVs with evolved donors %and periods of a few hours \citep{El-Badry_2021} are the natural progenitors of %extremely low-mass white dwarfs with more massive white dwarf companions %\citep{Aros-Bunster_2025,Belloni_2025}, which may then evolve into AM\,CVn %\citep{Belloni+Schreiber_2023}.

If the scenario just outlined is correct, the properties of CVs with evolved donors and longer orbital periods, corresponding to an earlier stage of the same evolutionary sequence, should be explicable under strong magnetic braking, as proposed by \citet{Van+Ivanova_2019}. Recently, \citet{Tovmassian_2025} showed that two long-period CVs, V479\,And and V1082\,Sgr, can indeed be explained as extremely long-period CVs with nuclear-evolved donors if such strong magnetic braking is assumed. Long-period CVs with evolved donors are not entirely new 
phenomena: GK\,Per (Nova Persei 1901), with an orbital period 
of $\sim$48\,h and an undermassive subgiant donor 
\citep{Alvarez-Hernandez_2021}, has long stood as an outstanding, but poorly understood example of this class. Recent massive surveys, however, have uncovered an increasing number of such systems, which we explore.

In this paper, we extend this investigation to SDSS\,J085210.48+783246.6 (hereinafter \obj), an eclipsing long-period CV whose fundamental 
parameters we derive and interpret within the same framework, 
showing that it is best understood as a progenitor of V479\,And 
at an earlier stage of the sequence.

\section{Comprehensive study of \obj}\label{sec:78}

\obj\ was identified as a CV candidate in a project that used machine learning on \textit{GALEX} ($FUV$, $NUV$) and \textit{Gaia} ($G$, $G_\mathrm{Bp}$, $G_\mathrm{Rp}$) absolute magnitudes plus an aggregate \textit{Gaia}-based variability metric (see Eq. 1 of \citealt{2021ApJ...912..125G}). We trained a Random Forest Classifier using 1200 spectroscopically confirmed CVs \citep{2003A&A...404..301R, 2011AJ....142..181S} as a training sample. Full details on the about 200 CV candidates identified in this project will be discussed elsewhere. 

The eclipsing nature of \obj\ and its very long orbital period of about 17\,h was determined from the analysis of the ZTF photometry \citep{Bellm_2019}, and its CV nature was confirmed by spectroscopy obtained at the Isaac Newton Telescope (see below).  {\sc Gaia} determines a parallax of $p=0.9180\pm0.0159$\,mas corresponding to a distance $D=1108.3^{+19.2}_{-18.5}$\,pc \citep{GAIA_DR3}.
\citet{Bailer-Jones_2021} calculates a slightly smaller distance of $1053_{-19}^{+17}$\,pc. Throughout this work, we adopt the \citet{Bailer-Jones_2021} distance, which incorporates a probabilistic prior on the Galactic stellar density distribution and is more robust than simple parallax inversion at this moderate signal-to-noise ratio.

\subsection{Observations}
We acquired spectrophotometric observations of \obj and complemented them with publicly available photometric data from ground-based and satellite telescopes.

\subsubsection{Spectrophotometry}

\obj\ was observed spectroscopically by the Sloan Digital Sky Survey V (SDSS-V) as part of its multi-epoch spectroscopic monitoring program. Three optical spectra were obtained using the BOSS spectrograph on the Apache Point Observatory 2.5\,m telescope \citep{Smee_2013}.
The BOSS spectrograph delivers a spectral resolution of $R \approx 2000$ over the wavelength range about 3600 to 10400\,\AA.

\begin{figure*}
\includegraphics[width=0.975\textwidth, bb = 0 0 700 410, clip=]{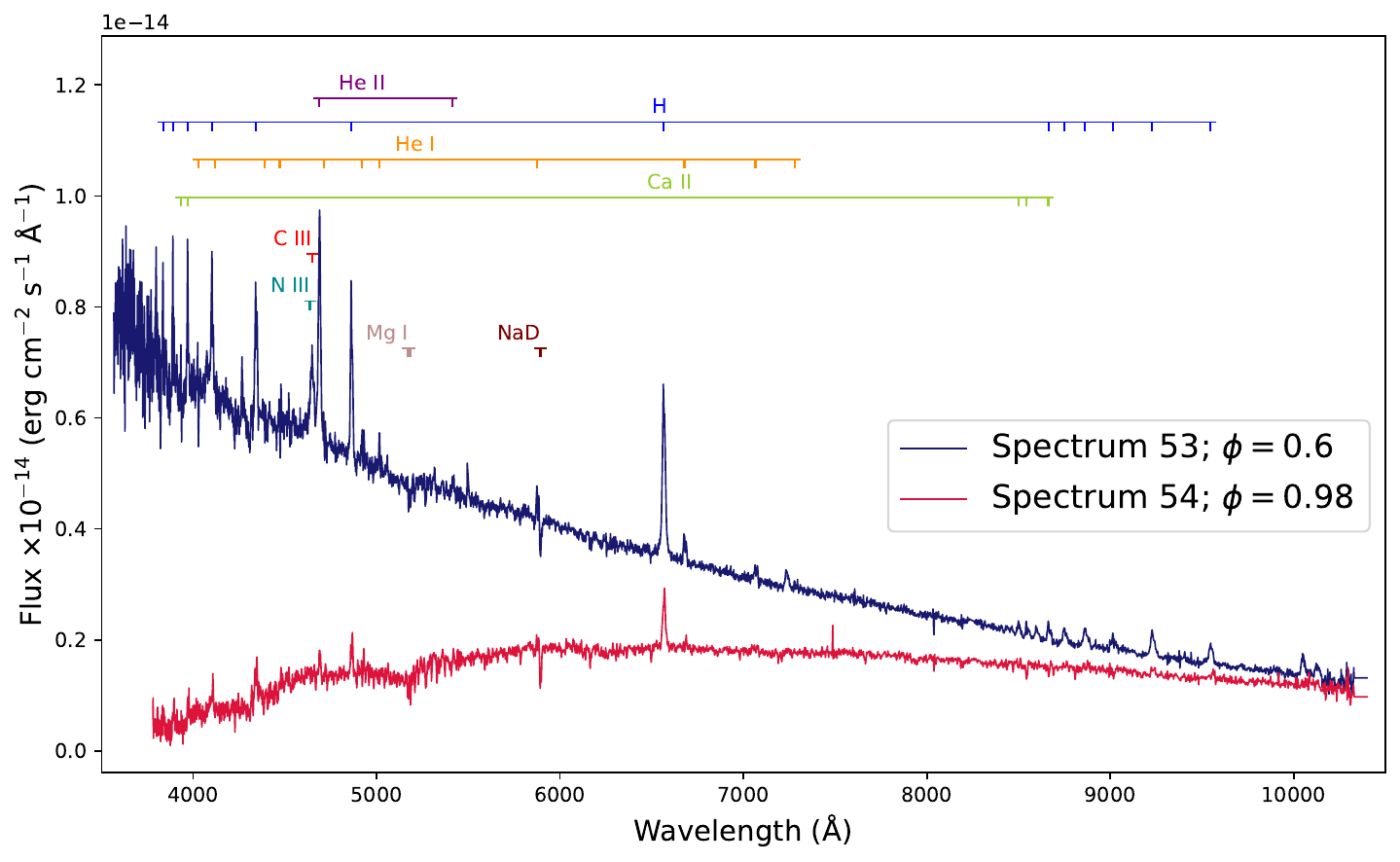}
\caption{SDSS BOSS spectra of SDSS J0852+7832 obtained during 
the bright phase (blue) and during eclipse (red). The spectra are 
shown in absolute flux units. The reduced flux level of spectrum \#54 
reflects its acquisition during eclipse.}
\label{fig:sloan}
\end{figure*}

The observations were carried out on Modified Julian Dates (MJDs) 59651, 59653, and 59654  which we denote as spectra \#51, \#53, and \#54, providing consecutive nightly coverage.
The spectra differ markedly from one another, and, as we found using the system ephemeris (see Section~\ref{sec:parameters}), spectrum \#54 was obtained during the eclipse and reveals prominent absorption features of the secondary star.
In contrast to \#54,  spectra \#53 and \#51 demonstrate strong Balmer, He\,\textsc{i}, and He\,\textsc{ii} emission lines typical for CVs. The spectrum \#53 was obtained at the maximum brightness of the system. In Fig.\,\ref{fig:sloan} we present spectra \#53 and \#54.
The major emission lines are marked. We omit the intermediary spectrum \#51 for clarity.

Phase-resolved observations of \obj\ were obtained
with the Intermediate Dispersion Spectrograph (IDS) mounted on the 2.54\,m Isaac Newton Telescope (INT) at the Observatorio del Roque de los Muchachos, La Palma. We used the R632V grating centred at 5720\,\AA, providing a spectral coverage from 4383 to 6704\,\AA\ at a dispersion of 0.89\,\AA\,pixel$^{-1}$. A total of 60 exposures of 600\,s each (10\,h total integration) were collected
between 2020 February 10 and 20. All spectra were optimally extracted and reduced using the \textsc{pamela} and \textsc{molly} packages \citep{Marsh1989}. Wavelength calibration was performed using Cu+Ne and Cu+Ar comparison lamps taken at the start and end of each night and approximately once per hour during the longer observing sequences. As a final refinement, we measured the position of the telluric O\,\textsc{i} $\lambda5577.4$\,\AA\ emission line in the sky spectrum and corrected any residual wavelength offsets to match its laboratory value. The INT spectra were not flux-calibrated.

Complementary observations were carried out with the 1.22\,m telescope at the Asiago Observatory equipped with the Boller \& Chivens spectrograph and an ANDOR iDus DU440A E2V 42-10 back-illuminated CCD. Observations took place between 2024 January 21 and 23, yielding fifty long-slit spectra evenly distributed over the orbital cycle. These spectra were obtained with a 1200\,l\,mm$^{-1}$ grating, providing a reciprocal dispersion of 0.6\,\AA\,pixel$^{-1}$ and a resolving power of about 2.3\,\AA\ (FWHM) over the 4580 {\bf to} 5820\,\AA\ range. An additional set of sixteen spectra was secured on 2024 February 1 using a 600\,l\,mm$^{-1}$ grating, covering 4250 {\bf to} 6590\,\AA\ with a 4.5\,\AA\ FWHM resolution.

All Asiago data were reduced following standard long-slit spectroscopy procedures in \textsc{IRAF}\footnote{\textsc{IRAF} (Image Reduction and Analysis Facility) is distributed by the National Optical Astronomy Observatory, which is operated by the Association of Universities for Research in Astronomy (AURA) under a cooperative agreement with the National Science Foundation.}, including bias subtraction, flat-field correction, and wavelength calibration using He+Ar arc-lamp exposures obtained contemporaneously with the science frames. Flux calibration was achieved with the spectrophotometric standard stars HR\,4554 and HR\,5501 observed on the same nights.

\begin{figure}
\centering
\includegraphics[width=0.47\textwidth, bb = 8 10 390 300, clip=]{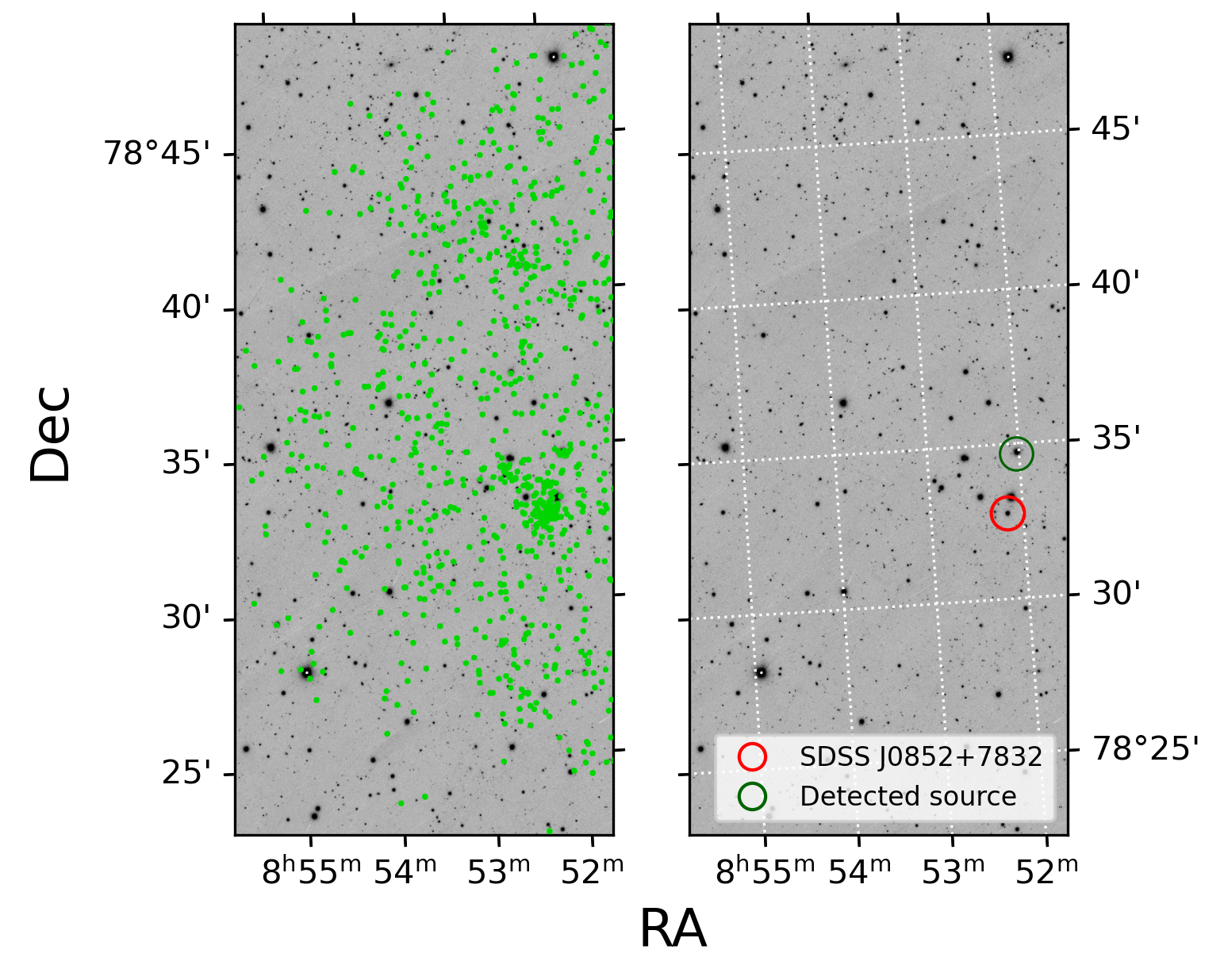}
\caption{The image of \obj\ in PanSTARRS $i-$band with overlayed {\sl Swift} XRT X-ray detection map (left panel). In the right panel, just the optical image is repeated with the \obj\ marked in addition to a randomly detected point source in this piece of sky.}
\label{fig:xray}
\end{figure}

\subsubsection{Photometry}

All-sky monitoring telescopes, including the All-Sky Automated Survey for Supernovae \citep[ASAS-SN;][]{Shappee_2014,Kochanek_2017}, Zwicky Transient Facility  \citep[ZTF;][]{Bellm_2019}, Asteroid Terrestrial-impact Last Alert System \citep[ATLAS;][]{Heinze_2018} and Transiting Exoplanet Survey Satellite \citet[TESS;][]{Ricker_2015} data have covered the object at different epochs and with a variety of filters. A summary of sky-patrol observations is presented in Table\,\ref{tab:observations}.

TESS observed the \obj\ on 2021-06-25 through 07-23, 2021-12-31 through 2022-01-07; 2022-06-13 through 07-08, and 2022-12-23 through 2023-01-18 with a standard cadence of 120\,s. We specifically used the light curves provided by the TESS Science Processing Operations Center \citep[SPOC\footnote{\url{https://archive.stsci.edu/hlsp/tess-spoc}}][]{Jenkins_2016}) through the Barbara A. Mikulski Archive for Space Telescopes (MAST\footnote{\url{https://archive.stsci.edu/missions-and-data/tess}}), and the PDCSAP flux, which corrects an initial simple aperture photometry (SAP) to remove instrumental trends and contributions from neighbouring stars using pre-search data conditioning. Nevertheless, examination of the TESS light curves shows negligible flux variability from epoch to epoch but more significant ({\bf about} 0.35\,mag) divergence at the bottom of the eclipse. The reason is not clear, but it can be mitigated by relying on multi-epoch, multi-colour observations from all-sky patrol missions, where no such large differences in the eclipse depth are observed.

\subsubsection{Swift UV and X-ray observations}

\swift\ \citep{Gehrels_2004} observed \obj\ upon our request (Target ID 165180). Ten exposures were distributed around the orbital phases to cover the entire orbital period. However, a gyroscope failure on board the spacecraft reduced the number of usable exposures; only four images of sufficient quality were obtained with the UVM2 filter on the UVOT telescope \citep{Roming_2005}. This is insufficient to analyse the light curve. However, we measured the average flux outside the eclipse. 

On the other hand, Swift/XRT \citep{burrows2000swift} shows a very low signal in each of the
$\approx 1000$\,s snapshots. In the combined $0.3$ to $10$\,keV image, our target appears as a
diffuse concentration of photons, distinguishable but not cleanly point-like. Fig.\,\ref{fig:xray}
displays the Pan-STARRS map \citep{Flewelling_2020} with the X-ray events overlaid; for
clarity, we repeat the optical image (right panel) and mark the positions of
SDSS~J0852+7832 and a nearby star (Pul-3~500095), the latter detected as a reasonably
strong X-ray source. At the optical position of SDSS~J0852+7832 we measure a PC-mode
count rate of $(3.7\pm0.9)\times10^{-3}$\,ct\,s$^{-1}$ in $0.2$ to $10$\,keV, extracted with a
standard 20-pixel ($ 47^{\prime\prime}$) circular aperture, which encloses 90\% of the
PSF at 1.5\,keV; the ARF from \texttt{xrtmkarf} applies the encircled-energy correction
\citep{Capalbi_2005}. The background was estimated from a concentric,
source-masked annulus with inner/outer radii of 40/80 pixels ($94^{\prime\prime}$ to $188^{\prime\prime}$),
with exposure-map weighting applied to both source and background regions to correct for
vignetting and bad pixels. For the count-to-flux conversion we adopt an absorbed thermal
bremsstrahlung model, \texttt{tbabs*bremss}, with $kT=7$\,keV and a Galactic column
$N_{\rm H}=2\times10^{20}$\,cm$^{-2}$ from the HI4PI map \citep{HI4PI_2016}; this choice is
appropriate for post-shock emission in accreting magnetic white dwarfs and reproduces the
observed hardness within the uncertainties. The resulting unabsorbed on-axis flux is
$(1.6\pm0.4)\times10^{-13}$\,erg\,cm$^{-2}$\,s$^{-1}$ in $0.2$–$10$\,keV. Although many XRT analyses
quote $0.3$ to $10$\,keV because of low-energy calibration systematics, the difference is
immaterial at such low $N_{\rm H}$ \citep{Godet_2009}. Adopting the distance used throughout
this work, $d=1053^{+17}_{-19}$\,pc, we obtain
$L_X=(2.12\pm0.52_{\rm stat}\pm0.07_{\rm dist})\times10^{31}$\,erg\,s$^{-1}$ via
$L=4\pi d^2 F$ \citep{Burrows_2005,Capalbi_2005,HI4PI_2016,UKSSDCXSELECT}. An additional systematic uncertainty of comparable magnitude arises from the assumed spectral model; this does not affect our conclusions.

\begin{figure}
\includegraphics[width=0.47\textwidth]{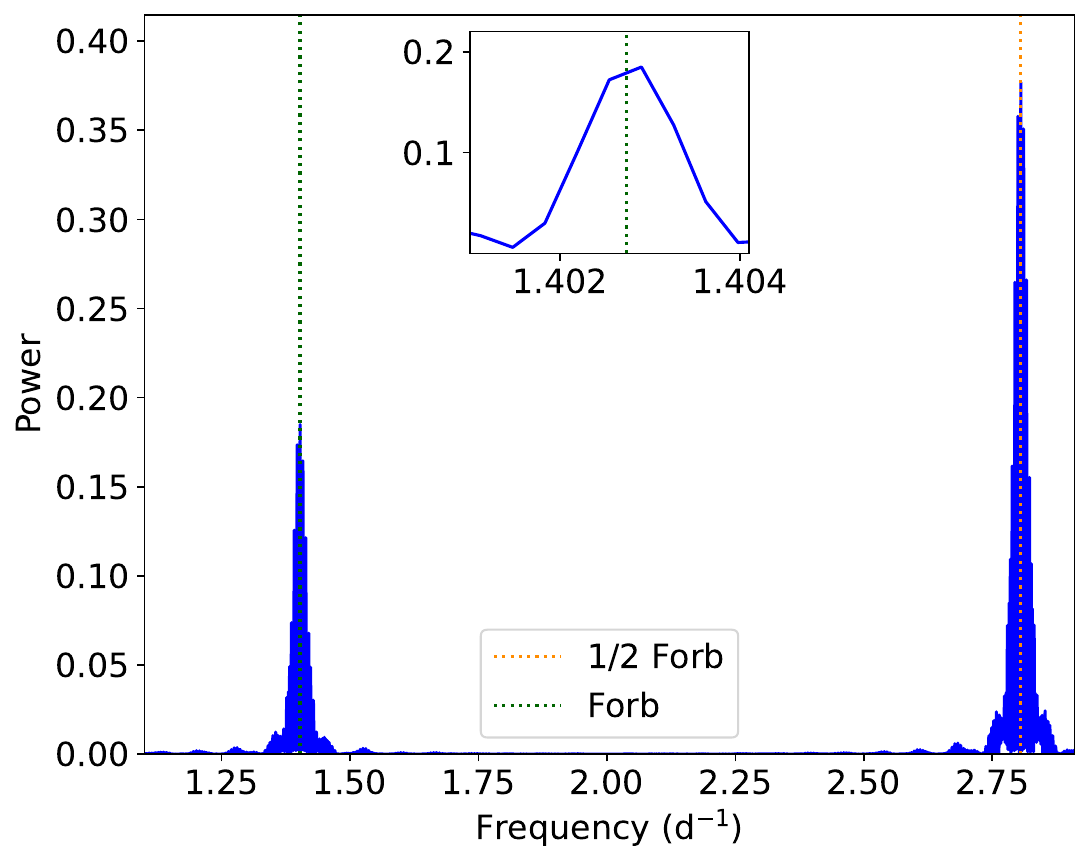}
\caption{The power spectrum of the combined photometric time series (TESS, ZTF, ASASSn, and Atlas). Two peaks correspond to frequencies of strong periodic variability corresponding to the orbital period and half of the orbital period, as a consequence of the double-humped light curve.The power is shown in normalized, dimensionless units.}
\label{fig:PowerSpectrum}
\end{figure}

\begin{figure}
\centering
\includegraphics[width=0.475\textwidth, bb = -5 23 580 465, clip=]{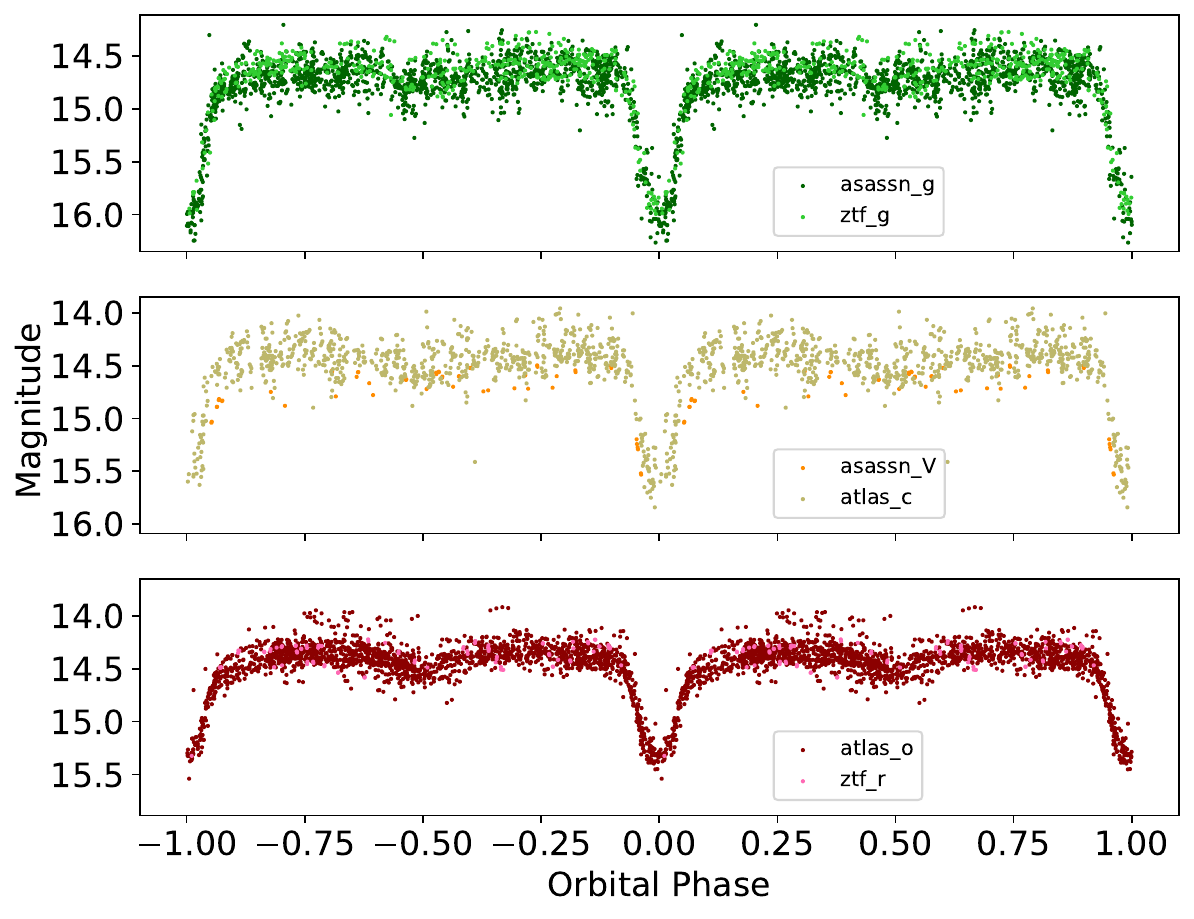}
\includegraphics[width=0.475\textwidth, bb = -5 0 690 325, clip=]{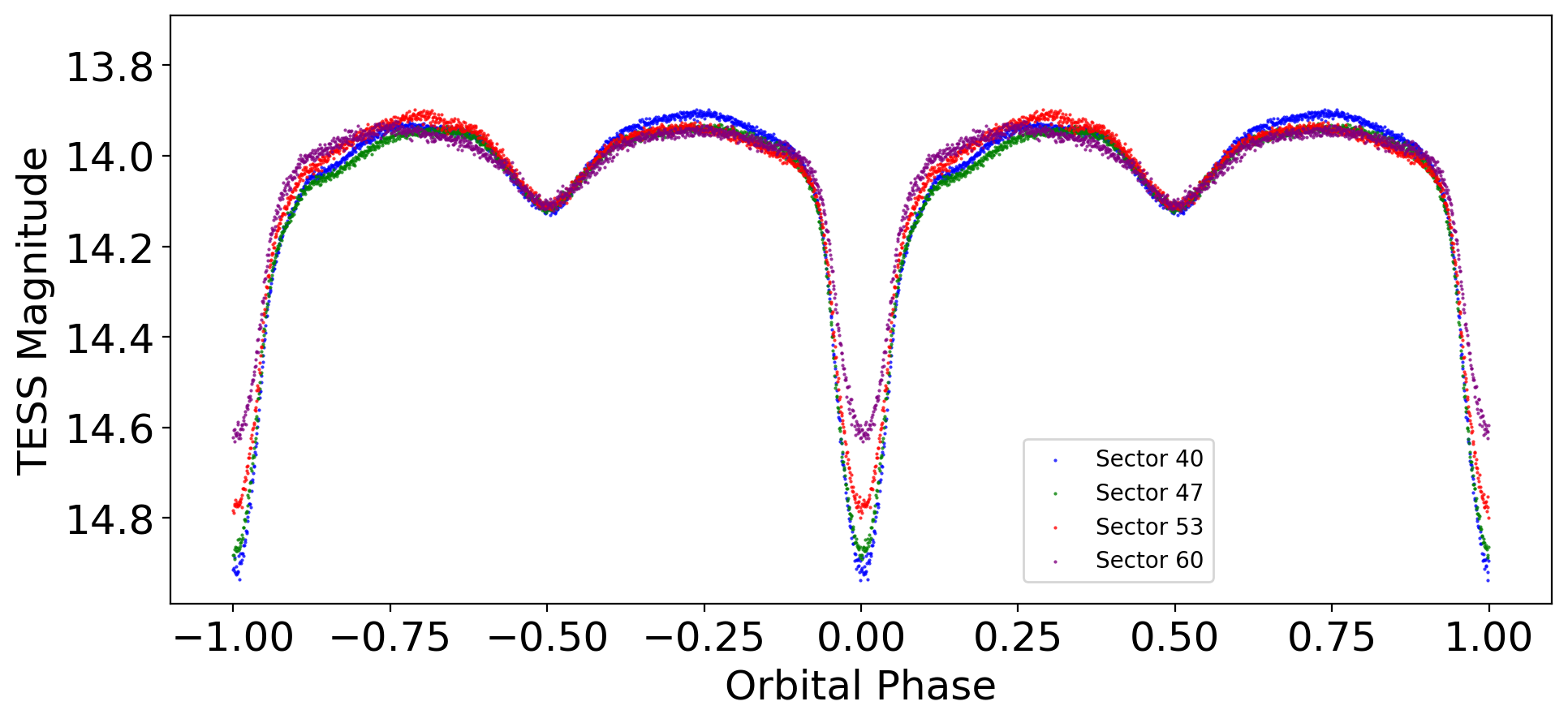}
\caption{The light curve of \obj\ in different bands collected by different sky patrol surveys (indicated in the legend) and folded with the orbital period is presented in the upper panel. In the lower panel, a binned TESS light curve is shown as points of different colours, corresponding to different epochs, marked by sector numbers in the legend. The overall brightness varies slightly from epoch to epoch; therefore, we have adjusted them to the same magnitude at phase 0.5 in Sector 53. While the shape of the TESS light curve is very definite, the depths of eclipses are not.}
\label{fig:lc}
\end{figure}

\section{System parameters} \label{sec:syspar}

From the observed photometric and spectroscopic data, we derive the 
fundamental parameters of the binary system. These parameters form the 
basis for the subsequent analysis of its physical properties and 
evolutionary status.

\subsection{Orbital period and ephemeris}
\label{sec:parameters}

A Lomb–Scargle periodogram (Fig.\,\ref{fig:PowerSpectrum}) computed from the detrended TESS light curve shows a dominant frequency at 1.4020(1)\,d$^{-1}$, corresponding to $P = 0.7129(1)\,\mathrm{d}$ (17.11\,h), which we identify as the orbital period. A secondary peak at twice this frequency reflects the ellipsoidal modulation of the donor star. The absence of significant aliases confirms that the period is unambiguously determined.

To refine the ephemeris, we measured the times of mid–eclipse by fitting Gaussian profiles to the minima in both the TESS and ground–based light curves, using data within orbital phases $-0.35$ to $+0.35$. This also allowed us to estimate the magnitude in the eclipse and its depth in different filters. All timestamps were converted to the barycentric dynamical time system (\(\mathrm{BJD}_{\mathrm{TDB}}\)). A linear least--squares fit to the resulting eclipse timings yields:

\begin{equation}
T_{\mathrm{mid}}(\mathrm{BJD}_{\mathrm{TDB}}) = 2458909.043(2) + 0.712892376(5)\,E,
\label{eq:ephem}
\end{equation}
where \(E\) is the orbital cycle number, \(T_0 = 2458909.043(2)\) corresponds to the time of mid-primary eclipse, and \Porb~$= 0.712892376(5)$\,d is the refined orbital period. No significant departure from a linear trend is detected in the observed minus calculated (O–C) residuals listed in Table~\ref{tab:observations}.
In Fig.\,\ref{fig:lc}, we present folded with the orbital period the available data from all surveys and TESS.

\begin{table*}
\centering
\caption{Eclipse depth, minimum magnitude, and O–C residuals for different surveys.}
\label{tab:observations}
\begin{tabular}{lccccc}
\hline
Survey & Eclipse depth (mag) & Min. magnitude & O–C (d) & No. Points & No. orbital cycles\\
\hline
ASAS-SN ($g$) & $-1.438$ & 16.13 &  0.000   & 1520 & 3193 \\
ZTF ($g$)     & $-1.718$ & 16.49 & $-0.002$ &   53 & 1227 \\
ASAS-SN ($V$) & $-1.189$ & 15.63 & $-0.003$ &  836 & 3473 \\
ATLAS ($c$)   & $-1.365$ & 15.99 & $-0.001$ &  466 & 4191 \\
ATLAS ($o$)   & $-0.996$ & 15.40 &  $0.000$ & 1795 & 4218 \\
ZTF ($r$)     & $-0.995$ & 15.35 &  $0.003$ &   89 & 1260 \\
\hline
\end{tabular}
\end{table*}

\subsection{Donor Star Spectrum Fitting and \texorpdfstring{$\chi^2$}{TEXT} Minimization}
\label{sec:dsparam}

\begin{figure*}
\setlength{\unitlength}{1mm}
\resizebox{15cm}{!}{
\begin{picture}(100,41)
\put (-12,0.5){\includegraphics[width=72.0mm, bb=0 10 720 420, clip=]{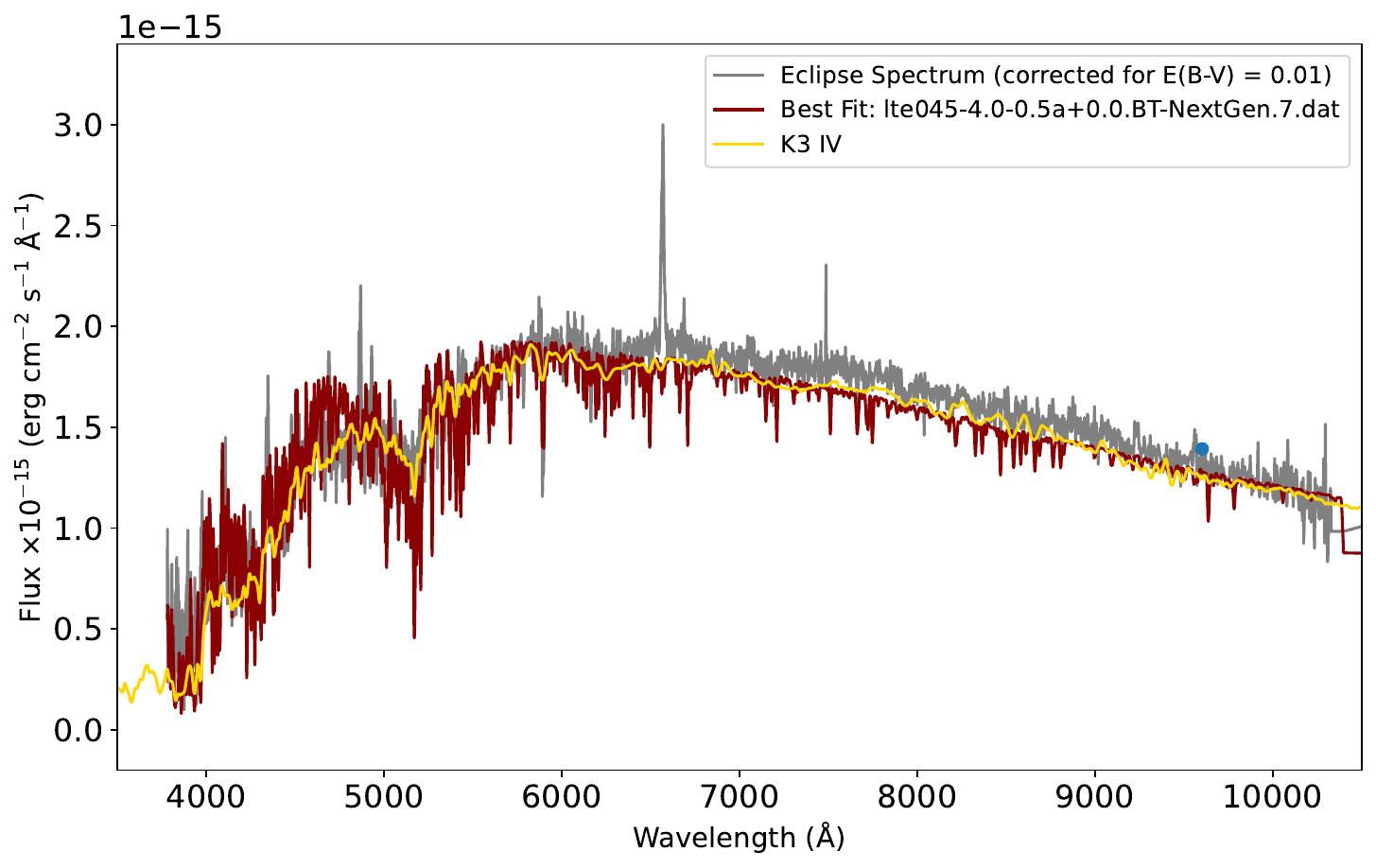}}
\put (57.5,1){\includegraphics[width=52mm, bb=0 8 560 450, clip=]{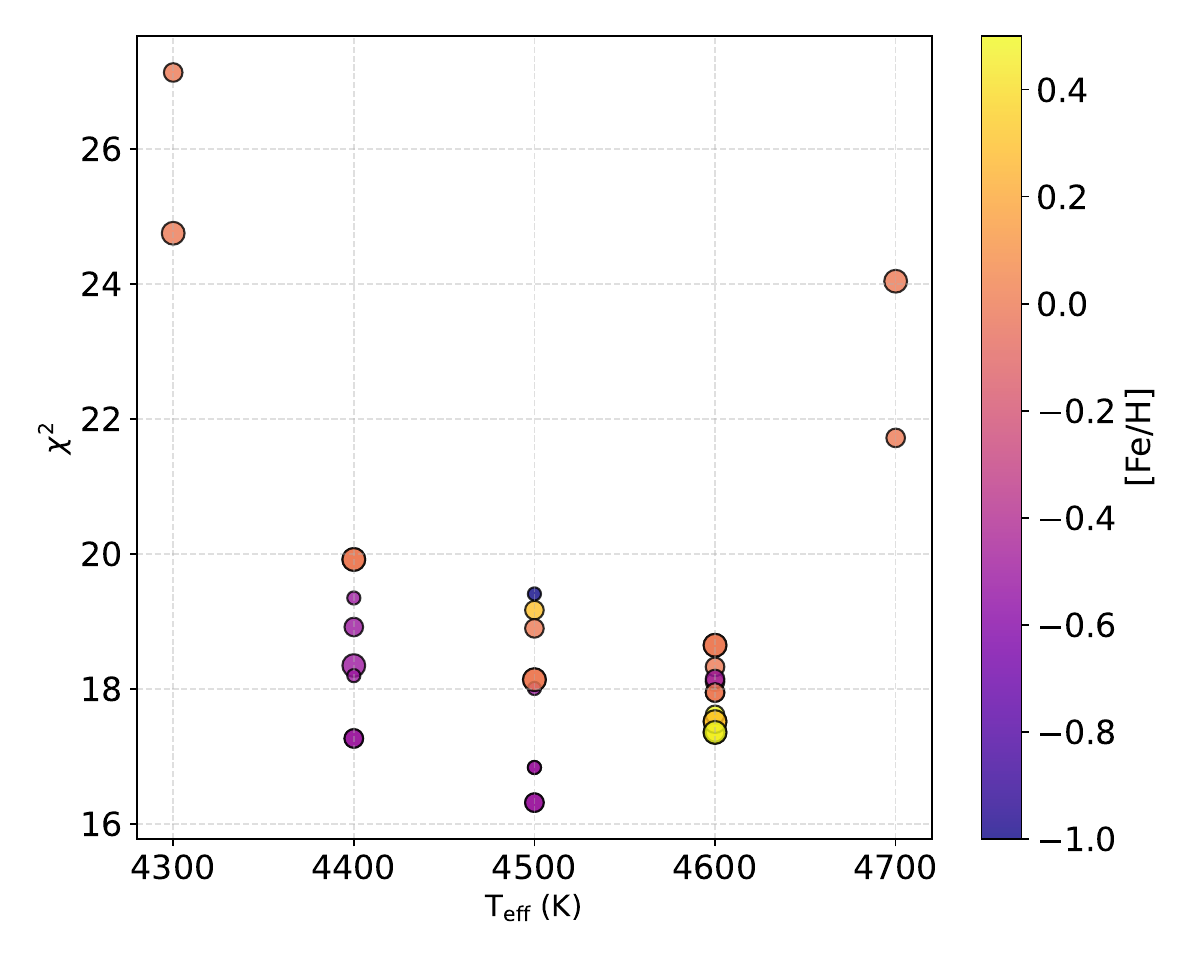}}
\end{picture}}
\caption{Left: Observed SDSS spectrum during eclipse, corrected for interstellar extinction (grey). The best-fit composite model (BT-Settl) is shown in dark red. The blue point indicates NIR photometric data; the rest are not shown in this figure but were included in the fitting. Right: Reduced $\chi^2$ for BT-Settl models in the grid considered during the fitting. Marker colour indicates effective temperatures, and size is scaled with $\log g$; larger circles correspond to $\log g\, (\mathrm{cm\ s^{-2}}) = 3.5$, smaller to $4.5$. A negative metallicity slightly improves the fit at lower temperatures, whereas a positive metallicity slightly improves it at higher temperatures. For solar abundances, the best fit is between 4500 and 4600\,K.
This diagnostic helps visualize the temperature--gravity region favoured by the observed spectrum. The formal best fit is achieved for \Teff $= 4500$\,K, $\log\,g (\mathrm{cm\ s^{-2}}) = 4.0$, $R=1.40$\,\Rsun, and $D=1073$\,pc.}
\label{fig:04}
\end{figure*}

To constrain the physical properties of the donor star in SDSS\,J0852+7832, we performed a detailed spectral energy distribution (SED) fitting using a grid of BT-Settl synthetic spectra\footnote{\url{https://phoenix.ens-lyon.fr/Grids/BT-Settl/AGSS2009/}} \citep{Allard_2014}. BT-Settl models incorporate detailed molecular opacities, dust formation, and convection, making them particularly well-suited for representing the atmospheres of cool stars with complex chemistry. The model grid we used spans a wide range of effective temperatures (4200–5000\,K, in 100\,K steps) and surface gravities ($3.5 < \log g \ (\mathrm{cm\ s^{-2}}) < 4.5$), which are appropriate for evolved K-type donors. The grid of synthetic spectra also contained models of different metallicities ([Fe/H]). They were compared to the observed optical spectrum 54 (it is centred at phase 0.9822 and covers 0.975 to 0.99 orbital phases) and near-infrared (IR) photometry corrected for interstellar reddening.\footnote{Interstellar extinction was corrected assuming $E(B-V) = 0.01$, using the recalibration of the \citet{Cardelli_1989} law by \citet{Fitzpatrick_1999}, via the \texttt{dust\_extinction} Python package.}

The observed data consist of two components,

\begin{itemize}
    \item an optical spectrum obtained during the white dwarf eclipse, providing an uncontaminated view of the donor star and
    \item near-IR photometric points in the $JHK$ bands, representing the donor's long-wavelength emission, where the contribution from the disc and the white dwarf is not significant. However, we must remember that the IR data available through \textit{Vizier} was most probably obtained outside of eclipse. Hence, it might exceed the IR tail of the SED in eclipse, simply because of the larger surface area of the RL-shaped donor star.
\end{itemize}

All model spectra were converted to physical flux units and scaled with the donor star's radius as a free parameter. Each synthetic spectrum was convolved to the resolution of the observed data and scaled by the geometrical dilution factor $(R/D)^2$, where $R$ is the radius of the donor and $D$ is the distance to the system. The grid spanned distances from 1033\,pc to 1073\,pc, bracketing the \citet{Bailer-Jones_2021} $1\sigma$ uncertainty interval, which is our adopted distance range---and radii from 1.2 to 1.6\,R$_{\odot}$. The raw \gaia\ parallax distance of $1108$\,pc lies outside this range but is disfavoured by the probabilistic distance prior (see Section~\ref{sec:78}).

\begin{figure}
\includegraphics[width=0.495\textwidth, bb = 0 0 700 440, clip=]{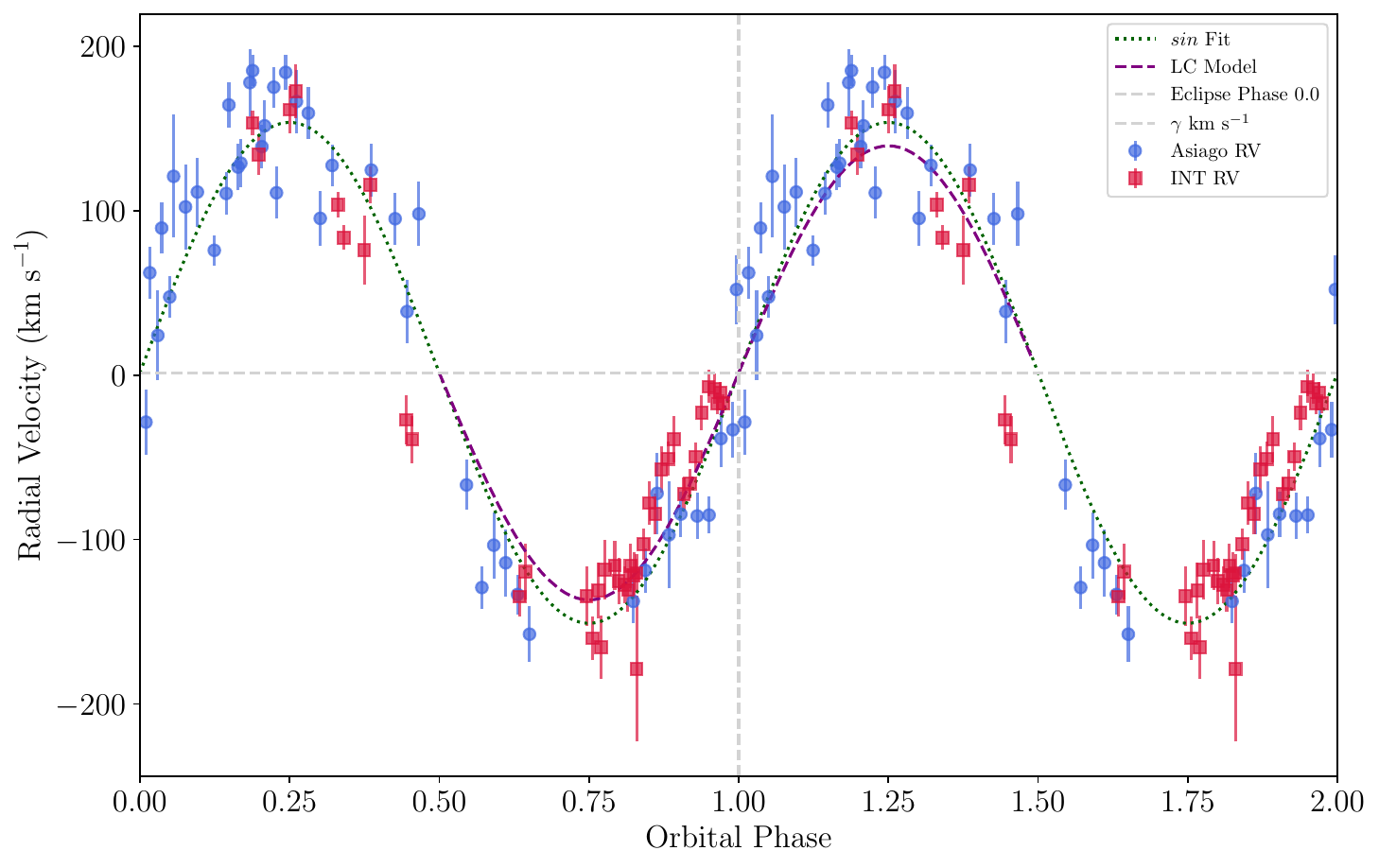}
\caption{The measured RVs of the donor star from both sets of data, Asiago and INT, as marked in the legend. Corresponding RV curves are fits of a sinusoidal function to the separate data sets. The ``model" RV curve was obtained by modelling the system parameters with light curves, as presented in Section\,\ref{sec:model}.}
\label{fig:RVsCurves}
\end{figure}

For each model, the reduced $\chi^2$ statistic was computed over the optical spectral range (3500 to 10500\,\AA), comparing the model to the ISM-corrected observed flux. The IR photometric points were excluded from the $\chi^2$ calculation because the archival IR data were most likely obtained outside eclipse and may reflect the larger projected emitting surface of the Roche-lobe-shaped donor; they were retained for qualitative consistency checks.

The best-fitting model corresponds to a BT-Settl atmosphere with $T_{\mathrm{eff}} = 4500$\,K, $\log g\, (\mathrm{cm\ s^{-2}})= 4.0$, and [Fe/H] = $-0.5$, scaled to a distance of 1053\,pc and a donor star radius of 1.40\,R$_{\odot}$. The resulting spectrum, shown in the left panel of Fig.\,\ref{fig:04}, provides the best match to the observed optical spectrum obtained during eclipse, when the contributions from the white dwarf and accretion disc are minimal, thus representing the donor star in its lowest observable emissivity state.

To validate the donor star spectral and luminosity class identification, we compare our best-fit model with a lower-resolution K3\,IV template spectrum from \citet[spectrum pickles\_uk\_59; $T_{\mathrm{eff}}= 4570$\,K,][]{Pickles_1998}, shown in the left panel of Fig.\,\ref{fig:04}.

To explore the fit sensitivity to stellar parameters and potential degeneracies, we plotted $\chi^2$ as a function of effective temperature (right panel of Fig\,~\ref{fig:04}). Each point in the diagram corresponds to a model with a specific $T_{\mathrm{eff}}$, $\log g$, and [Fe/H]. Points are colour-coded by metallicity and sized according to surface gravity (larger symbols represent lower $\log g$). The diagram reveals a clear minimum around $T_{\mathrm{eff}} = 4500$\,K and $\log g (\mathrm{cm\ s^{-2}}) \approx 4.0$, with acceptable fits extending within $\pm100$\,K of the minimum.

This visualization not only identifies the best fit but also provides insight into the structure of the parameter space, illustrating how temperature and metallicity interact to shape the observed spectrum.

The apparent numerical differences between the temperatures derived
from SED fitting ($T_{\rm eff} = 4500$\,K) and the diagnostic
diagram ($T_{\rm eff} = 4580$\,K) reflect the finite resolution of
the BT-Settl grid (100\,K steps in $T_{\rm eff}$, 0.5\,dex steps
in $\log g$) rather than a genuine physical disagreement; any temperature
in the range $4500$ to $4650$\,K is an equally acceptable fit to the
observed spectrum. The surface gravity is similarly constrained only
to within the grid interval, and the best-fit $\log g\,(\mathrm{cm\ s^{-2}}) = 4.0$ should
be understood as indicating an evolved, subgiant-like atmosphere
rather than a precise measurement. Regarding metallicity, the
available BT-Settl spectra cover only a discrete set of [Fe/H], so the best-fitting [Fe/H]\,$= -0.5$ is indicative rather
than a definitive abundance determination; a proper measurement
would require higher-resolution spectroscopy and a dedicated
abundance analysis. We note that the MESA evolutionary models in
Section~\ref{sec:evol} assume $Z=0.015$ (near-solar metallicity),
which is required to reproduce the observed donor temperature and
radius. This is nominally higher than the sub-solar metallicity
suggested by the spectral fitting, but given the coarse [Fe/H]
resolution of the BT-Settl grid and the absence of a dedicated
abundance analysis, we do not consider this a significant
inconsistency.

Using a complex of absorption lines in the $\lambda\lambda \ 5100$–$5500$\,\AA\ range and a standard star template \citep[HD165341, K0\,V;][]{Prugniel_2001}, we measured the radial velocities (RVs) of the donor star from both the Asiago and INT observations. The RVs were determined by a cross-correlation method\footnote{We used the \texttt{xcsao} task from the RVSAO package within IRAF \citep{Kurtz_1998}, which performs Fourier cross-correlation between object and template spectra.}. The orbital period derived from the RV variations agrees, within the uncertainties, with the photometric period, the latter being significantly more precise. The phase-folded RV curves are shown in Fig.\,\ref{fig:RVsCurves}. A slight difference in the amplitude of the sinusoidal fits is observed between the two data sets. The combined result yields $K_{\mathrm{d}} = 151 \pm 10$\,\kms.

To constrain the donor mass, we employ the $R_2$--$M_2$ diagnostic diagram
introduced by \citet{Tovmassian_2025}. The method combines the observed
orbital period with the assumption that the donor star fills its Roche
lobe. For a binary with a given orbital period, the Roche-lobe radius is
uniquely determined by the masses of the two components through the
binary separation and the mass ratio. Consequently, for each assumed
white dwarf mass, one can calculate the Roche-lobe size as a function of
the donor mass.

\begin{figure}
\centering
\includegraphics[width=0.495\textwidth, bb = 0 5 530 450,clip=]{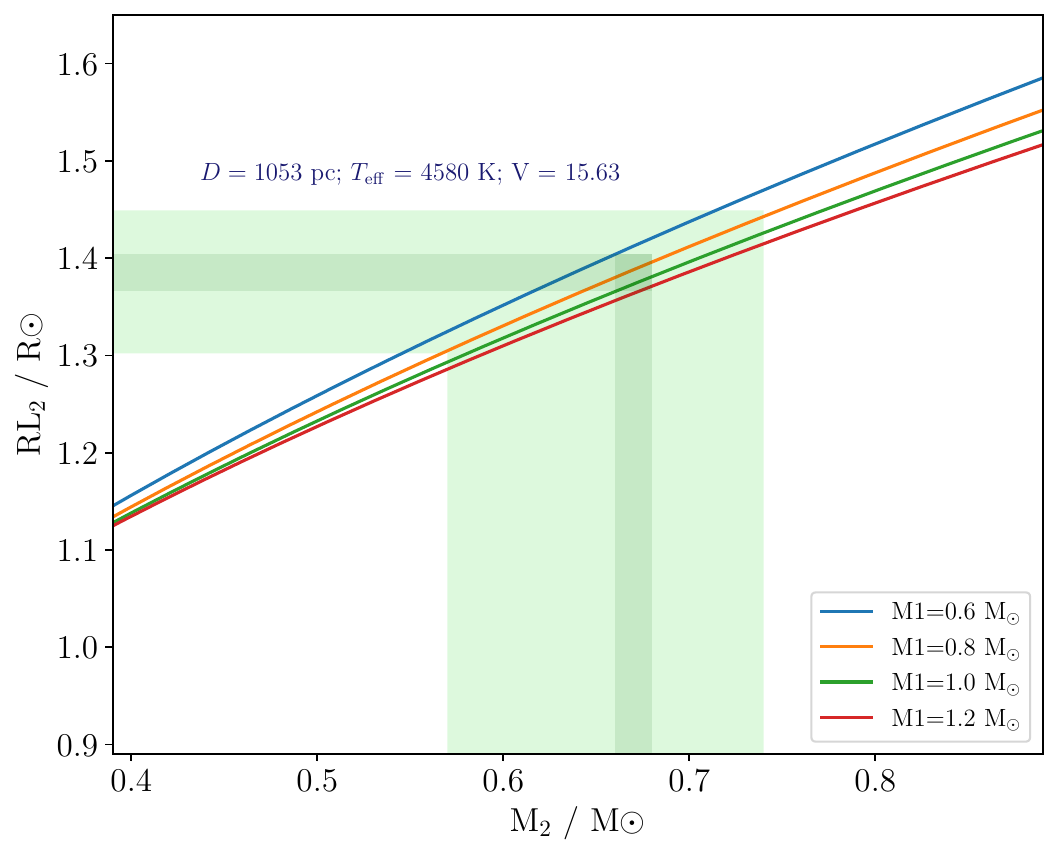}
\caption{A diagnostic diagram of $ R_2-M_2$ relation for \obj \ assuming that the donor star fills its Roche lobe and a spherical geometry. The light shaded area indicates the range of calculated radii of the donor stars in a full range of Gaia distances and the uncertainty of brightness estimates in the eclipse. The darker shaded area corresponds to T$_{\mathrm{eff}}=4580$\,K, yielding the observed luminosity of the donor star for the most probable distance of 1053 pc. A bolometric correction BC$=-0.395$ and interstellar extinction E$(B-V)=0.01$ were applied to the estimated $m_{V}=15.63\pm0.1$ at the bottom of the eclipse to convert to the luminosity. The intersection of dark shaded areas indicates the preferred solution.}
\label{fig:DIagnosticDiagram}
\end{figure}

This produces a family of curves in the $R_2$--$M_2$ plane (Fig.\,\ref{fig:DIagnosticDiagram} representing
the radius that a Roche-lobe filling donor must have in a binary with the observed orbital period. Each curve corresponds to a different assumed white dwarf mass. The donor radius can be estimated independently from its brightness and effective temperature, with the Gaia distance.
The Roche lobe filling donor star has a certain luminosity\footnote{A bolometric correction BC$=-0.395$ was applied according to \citet{Montegriffo_1998,Eker_2020}} for the distance determined by \gaia \ D$=1053^{+16}_{-19}$ and given radius and $T_{\mathrm{eff}}$ = 4580 K. These observational constraints appear in the diagram as a horizontal band corresponding to the allowed range of $R_2$. The light-shaded region in Fig.\,\ref{fig:DIagnosticDiagram} encompasses the full range of donor radii consistent with the $1\sigma$ Gaia distance uncertainty (1033 to 1127\,pc), and uncertainties related to the temperature and brightness of the donor star and its luminosity estimate.

The intersection between the observational radius constraint and the Roche-lobe curves provides the corresponding donor mass for each
assumed white dwarf mass. In this way, the diagram allows us to visualize the allowed region of the $R_2$--$M_2$ parameter space and to determine the donor mass largely independently of the details of the light-curve modelling.
 According to Fig.\,\ref{fig:DIagnosticDiagram} the donor mass is $\approx0.68^{+0.03}_{-0.07}$ \Msun. The tendency is toward lower mass, because the eclipse magnitude from ASAS-SN($V$) is probably overestimated (ASAS-SN$(V) \approx $ ASAS-SN$(g) -0.3$).
Hence, the donor remains significantly undermassive relative to an
isolated K3--K4 main-sequence star \citep{Boyajian_2012} across
this entire range. The main conclusion is therefore robust against
the distance uncertainty.

\subsection{Light curve model}\label{sec:model}

\begin{figure*}
\centering
\includegraphics[width=0.475\textwidth, bb = 0 0 720 620, clip=]{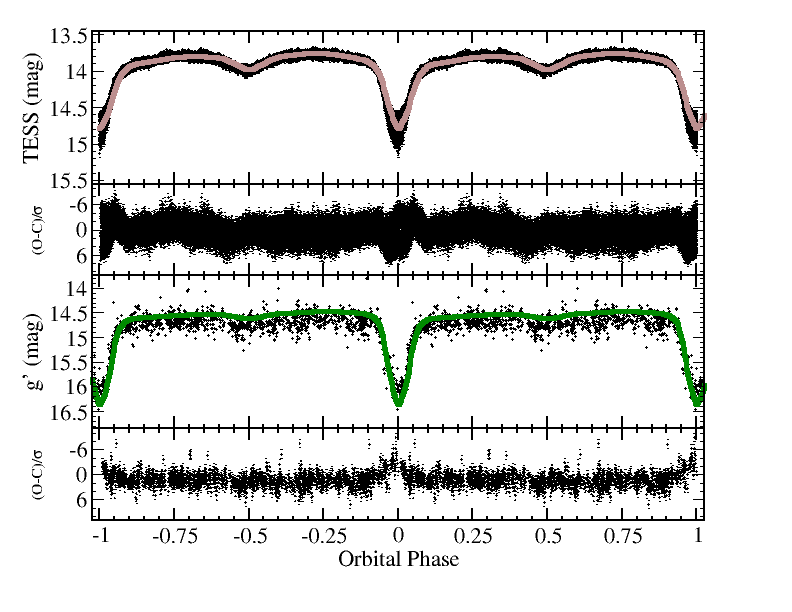}
\includegraphics[width=0.475 \textwidth, bb = 0 0 720 620, clip=]{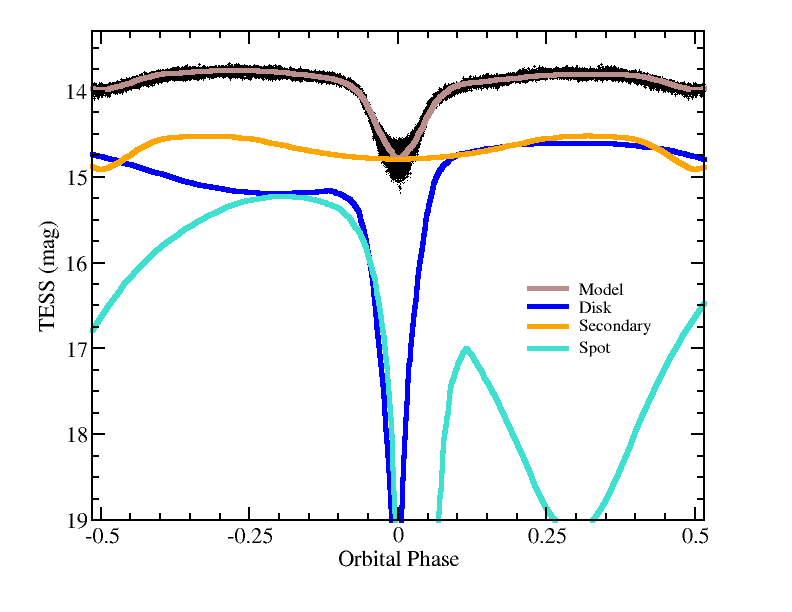}
\caption{Light curve model fitting. In the left panel, the $g'$-band light curve of \obj\ is shown, along with the best-fit model in the upper panel. In the bottom panel, (O-C)/$\sigma$ for data vs.\ model. In the right panel, the contributions of all components of the binary are presented to produce the modelled light curve.}
\label{fig:05ab}
\end{figure*}

\begin{figure}
\centering
\includegraphics[width=0.475\textwidth,bb = 50 40 1450 1325, clip=]{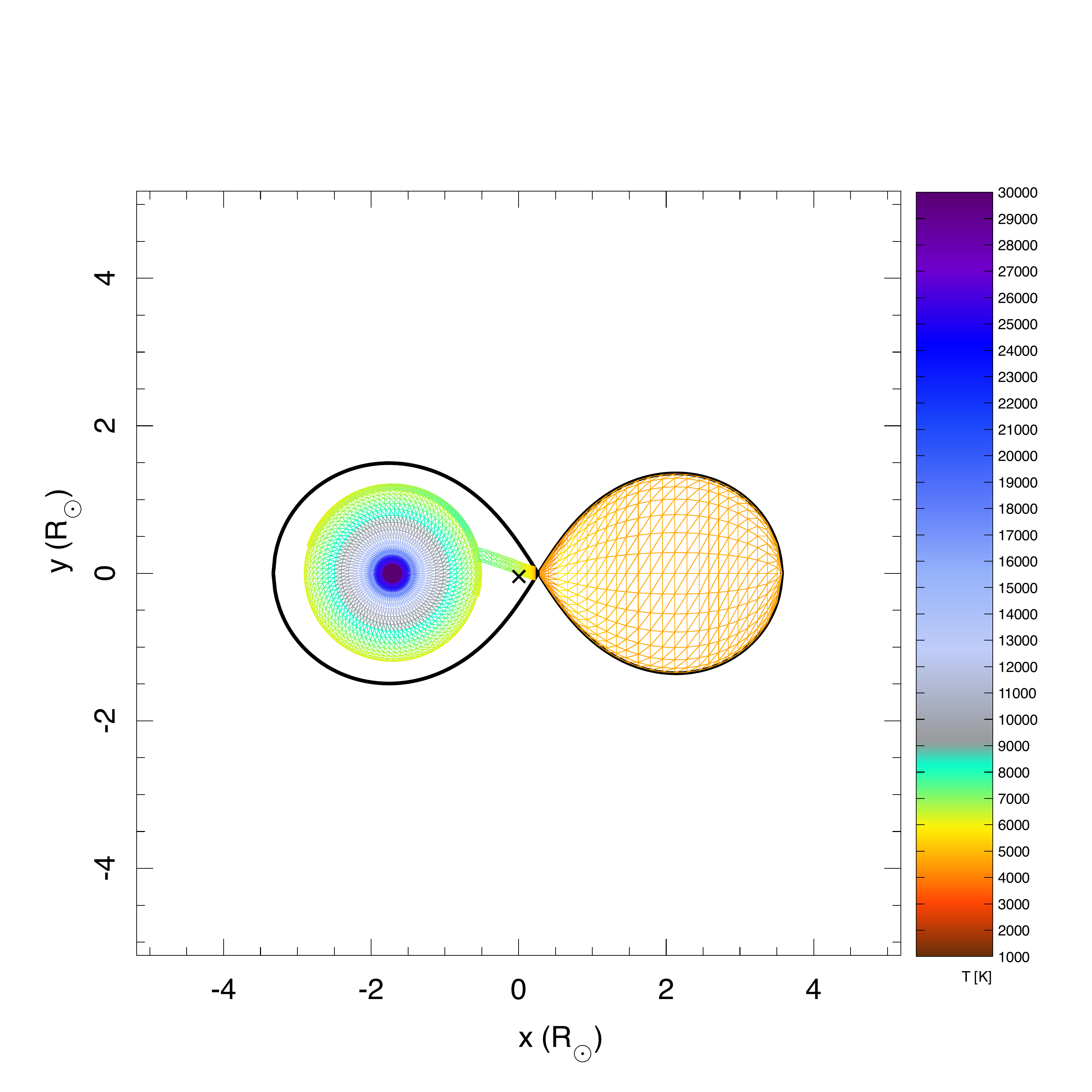}
\caption{The composition of \obj\ corresponding to the best fit of the light curve. The axes show the dimensions of the components of the binary system in solar radii. Symbol ``x'' corresponds to the centre of mass.}
\label{fig:05}
\end{figure}

We used the $g'$-band and TESS light curve (LC) to apply a modelling tool developed by \citet{Zharikov_2013}, and described in detail in \citet{Subebekova_2020} and \citet{Kara_2021, Kara_2023}, to determine the system parameters and to study the structure of the accretion disc. The model consists of four components: the white dwarf, the donor star, the accretion disc, and the bright spot. Each component is represented by a set of triangles (projection) with certain $T_{\mathrm{eff}}$ projected along the line of sight.
The model assumes that the disc radiates as a black body at the local effective temperature with radial distribution across the disc given by:

\begin{equation}
\begin{aligned}
T_{\rm eff}(r)   &= T_0 \left\{ \left(\frac{r}{R_{\rm WD}}\right)^{-3} \, \left(1-
\left[\frac{R_{\rm WD}}{r}\right]^{1/2}\right)\right\}^{EXP},  \\
 T_0  &=  \left[ \frac{3G\,M_{\rm WD}\,\dot{M}}{8\pi\sigma\, R_{\rm WD}^3} \right]^{1/4},
 \end{aligned}
\label{Tempeq}
\end{equation}

\noindent
where
$R_{\mathrm{WD}}$ is the radius of the non-rotating helium WD \citep{Nauenberg_1972},
$\dot{M}$ is the mass transfer rate, and $\sigma$ is the Stefan--Boltzmann constant. 
In the standard accretion disc model, the radial temperature 
gradient is taken to be $\mathrm{EXP} = 0.25$  \citep[equation 2.35]{Warner_1995}.
We allow it to deviate slightly from this value, as in \citet{Linnell_2010}. 
The accretion disc thickness is defined as
\begin{equation}
z_\mathrm{d}(r) = z_\mathrm{d}(r_{\mathrm{out}})(r/r_{\mathrm{out}})^{\gamma_{\mathrm{disc}}},
\end{equation}
where $\gamma_{\mathrm{disc}}$ is a free parameter for which we used the standard value of $\gamma_{\mathrm{disc}}= 9/8$ \citep[equation 2.51b]{Warner_1995} as the initial value. We also used the complex model for the hot spot, as described in detail by \citet[see their Fig.\,9 and therein]{Kara_2021}, where all parameter definitions can be found. The disc limb-darkening used in the model follows the Eddington approximation \citep{Mayo_1980, Paczynski_1980}. The gravitational potential within the corresponding Roche lobe determines the shape of the stellar surface. The code includes the quadratic limb-darkening law for the primary and secondary with the limb-darkening coefficients taken from \citep{2012A&A...546A..14C, 2013A&A...552A..16C, 2020A&A...634A..93C}.

We fit the light curve by minimizing the $\chi^2$ function, setting all parameters free, and using the gradient descent method to obtain self-consistent fit results. The best-fit parameters are given in Table~\ref{tab:BestPar} and the resulting light curves and the contribution of each component are presented in Fig.\,\ref{fig:05ab}, left and right, respectively. The numbers in brackets for the variables in Table~\ref{tab:BestPar} are uncertainties defined as 1$\sigma$ of the Gaussian function approximation, which is used to describe the one-dimensional $\chi^2$ function. The model predicts that in the $g'$-band, the light curve is formed by combining flux from two main components: the secondary and the disc. The contribution to the continuum from the hot spot and WD is significantly lower than theirs. The size of the accretion disc is expected to be the tidal truncation radius. For the derived mass ratio $q=0.82$, the tidal truncation radius is approximately $0.35$--$0.40$ times the binary separation \citep[e.g.][]{Warner_1995}, giving $R_{\rm tid}\simeq1.3$--$1.5$\,\Rsun\ for $a=3.78$\,\Rsun, consistent with the fitted value of $R_{\rm d,out}=1.20^{+0.05}_{-0.2}$\,\Rsun.

The geometry of the system is shown in Fig.\,\ref{fig:05} for a better understanding of the system and disc parameters. The colour bar in the figure indicates the effective temperature of the black body, corresponding to radiation from the system components.

\begin{table}
\centering
\caption{System parameters from the lightcurve modelling}
\label{tab:BestPar}
\begin{tabular}{llll}
\hline\noalign{\smallskip}
{\bf Fixed parameters:}       &            &          \\
\hline\hline\noalign{\smallskip}
\multicolumn{2}{l}{$P_{\mathrm{orb}}$}  & \multicolumn{2}{c} {17.109 h}   \\
\multicolumn{2}{l}{$E(B-V)$ }           & \multicolumn{2}{c} {0.01 mag.}    \\
\multicolumn{2}{l}{Distance}            & \multicolumn{2}{c} {1053.1 pc}  \\ \multicolumn{2}{l}{${T}_{\mathrm{WD}}$} & \multicolumn{2}{c}{30000 K}\\

\hline\noalign{\smallskip}
\multicolumn{4}{l}{{\bf Variables of the LC fit:}} \\  \hline\noalign{\smallskip}
\hline\noalign{\smallskip}
\multicolumn{4}{l}{{ Parameters of the system:}} \\
\hline\noalign{\smallskip}
\multicolumn{2}{l}{ $i$ }               & \multicolumn{2}{c}{81.5$^\circ$(1.0) } \\
\multicolumn{2}{l}{$M_{\mathrm{WD}} $ } &\multicolumn{2}{c}{0.80(4) M$_{\sun}$ }  \\
\multicolumn{2}{l}{$q = M_2/M_{\rm WD} $}  & \multicolumn{2}{c}{ 0.82(9)}  \\
\multicolumn{2}{l}{$T_{2, night}$ }  & \multicolumn{2}{c}{$ 4700(100)$  K}\\
\multicolumn{2}{l}{$T_{2, day, max, L_1}$ }  & \multicolumn{2}{c}{$ 5800^{+300}_{-600}$  K}\\
\multicolumn{2}{l}{$\dot{M}$$\times10^{-8}$\Msunyr} & \multicolumn{2}{c}{7.1(8)}\\

\hline\noalign{\smallskip}
\multicolumn{4}{l}{{ Parameters of the disc:}} \\
\hline\noalign{\smallskip}
\multicolumn{2}{l}{ $R_{\mathrm{d, in}}$  } & \multicolumn{2}{c}{$\equiv R_{\rm WD}$ } \\
\multicolumn{2}{l}{ $R_{\mathrm{d, out}}$ } & \multicolumn{2}{c}{ 1.20$^{+0.05}_{-0.2}$ R$_{\sun}$ } \\
\multicolumn{2}{l}{ $h_{\mathrm{d, out}}$ } & \multicolumn{2}{c}{ 0.150$^{+0.025}_{-0.015}$  R$_{\sun}$ } \\

\noalign{\smallskip}\hline\noalign{\smallskip}
\multicolumn{4}{l}{{\bf Calculated system parameters$^*$:}} \\
\hline\noalign{\smallskip}
\multicolumn{2}{l}{$a $}  &\multicolumn{2}{c}{3.78  R$_{\sun}$}  \\
\multicolumn{2}{l}{$M_{2}$}  & \multicolumn{2}{c}{0.62 M$_{\sun}$} \\
\multicolumn{2}{l}{$R_{2,x}$}  & \multicolumn{2}{c}{1.38 R$_{\sun}$}\\
\multicolumn{2}{l}{$R_{2,y}$}  & \multicolumn{2}{c}{1.79 R$_{\sun}$}\\
\multicolumn{2}{l}{$\log g_{2}$}  & \multicolumn{2}{c}{3.97}\\
\multicolumn{2}{l}{$R_{\mathrm{WD}}$ } & \multicolumn{2}{c}{0.01 R$_{\sun}$}\\
\multicolumn{2}{l}{$\log g_{1}$}  & \multicolumn{2}{c}{8.36}\\
\hline \\
\end{tabular}
\begin{tabular}{l}
$^*$ Those parameters correspond to the best fit. \\
Numbers in brackets are uncertainties of variables.
\end{tabular}
\end{table}

\subsection{Overall SED and the composition of the binary}

\begin{figure}
\centering
\includegraphics[width=0.45\textwidth]{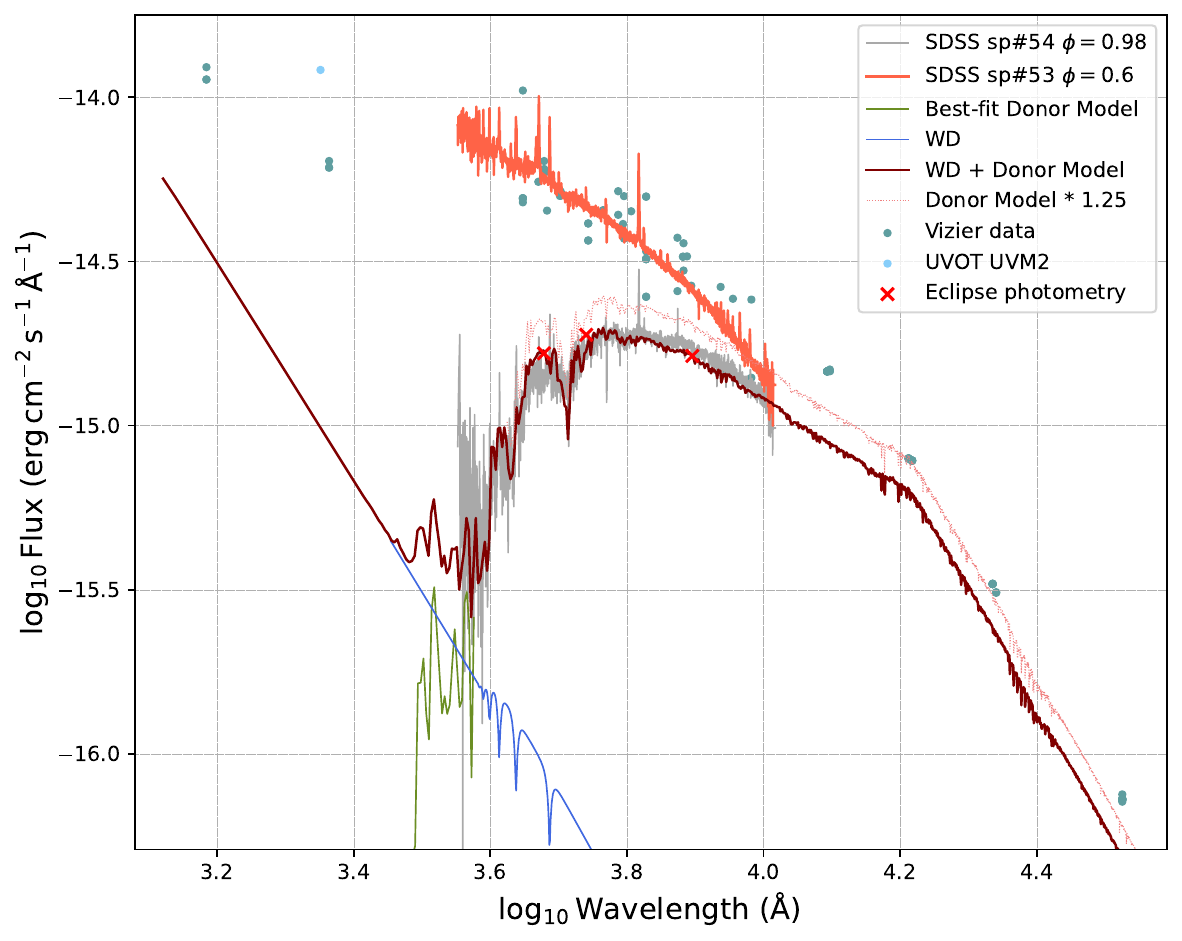}
\caption{The SED of the \obj\ comprised of SDSS-V spectra taken in two epochs, one in the eclipse, the other around the maximum. Overplotted is the best-fit donor star model obtained from the grid of BT-Settl model atmospheres for Stars \citep{Allard_2014}, as in Fig.\,\ref{fig:DIagnosticDiagram}, but with a downgraded resolution for better illustration. To achieve a better fit of IR points (the dotted green line), it is sufficient to increase the radiating surface by a factor 1.25, which agrees with the Roche lobe projection ratio of $R_{\mathrm{2,y}}/R_{\mathrm{2,x}}$ (see Table\,\ref{tab:BestPar}). We used a WD model from \citep{Koester_2010} of $T=30{,}000$\,K and log g=8.335 (see text). The composite spectrum of the two stellar components is also shown, assuming that the contribution of the accretion disc to the continuum during the eclipse is negligible. In addition, the Vizier photometry of the object and the \swift\ (UVOT) measurement in the UVM2 band are plotted as points, all referenced in the legend. The fluxes in eclipse determined from the ASSASN are marked by crosses.}
\label{fig:SED}
\end{figure}

In addition to optical spectral fitting, we constructed the full SED of the system, spanning from the UV to the mid-IR, and presented it in Fig.\,\ref{fig:SED}. The SED includes archival GALEX fluxes, our own Swift UVM2 measurement, optical spectroscopy during eclipse, and near- to mid-IR photometry. The plot also features the \SDSSV\ spectrum \#53, obtained at the brightest orbital phase of 0.6. Using the best-fit parameters for the donor star ($T_{\mathrm{eff}} = 4500$\,K, $\log g = 4.0$), and the white dwarf parameters derived from light curve modelling ($T_{\mathrm{eff}} = 30{,}000$\,K, consistent with the value adopted in Table~\ref{tab:BestPar}, $R_{\mathrm{WD}} = 0.0099$\,R$_{\odot}$), we generated a composite SED to compare it with the observed data. An unaccounted-for contribution from the accretion disc naturally explains the difference between the composite model and observations in the UV.

The difference between the observed and model flux at the $H$ band is approximately 20\%. We do not know the orbital phases at which the IR data were obtained, but they were likely obtained outside eclipse. This discrepancy is due to the projected emitting surface of the donor star being much larger toward the back (toward the observer during eclipse) and toward its side, which qualitatively explains the observed infrared excess. Hence, multiplying the best-fit model by 1.25 improves the model's fit to the IR data.

\section{Emission lines and characteristics of the accretion disc}

\begin{figure}
\setlength{\unitlength}{1mm}
\resizebox{10cm}{!}{
\begin{picture}(100,94)
\put (0,0){\includegraphics[width=11cm, bb= 150 220 635 650, clip=]{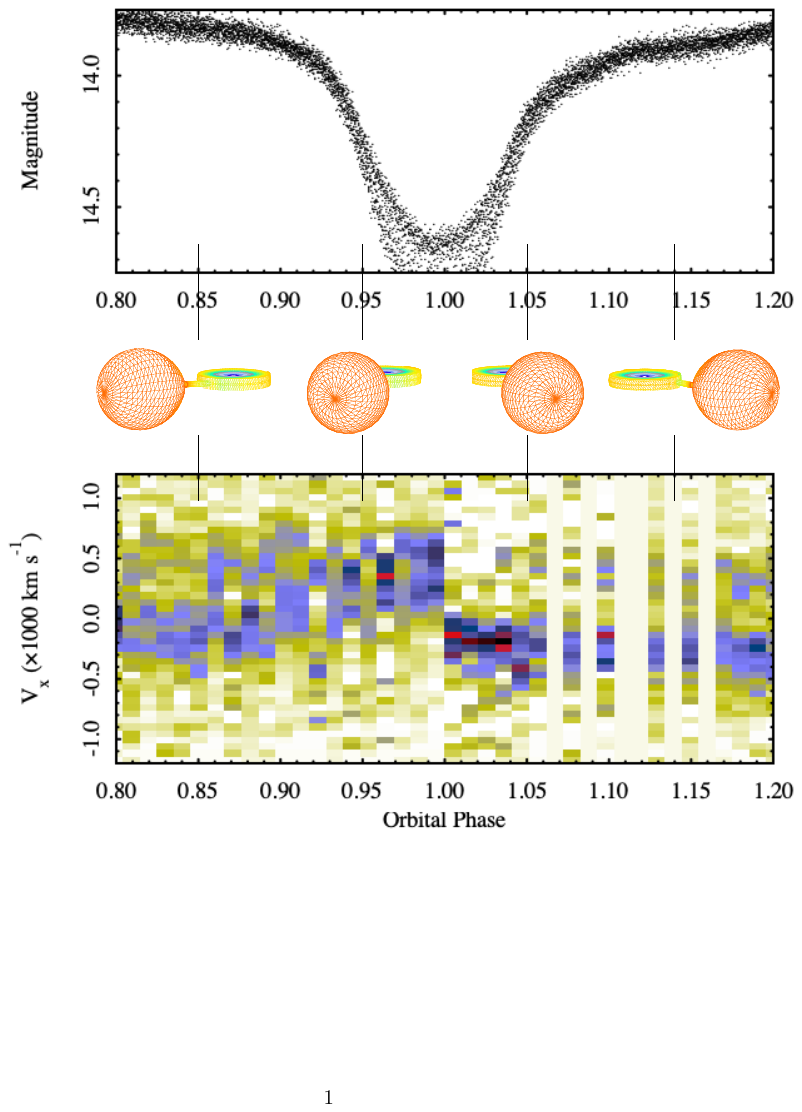}}
\end{picture}}
\caption{The eclipse morphology: In the upper panel, a portion of the TESS light curve around the eclipse is presented to be compared with the H$\beta$ emission line behavior displayed in the form of a trailed spectrum in the bottom panel. An ingress and egress of the disc are embodied as a jump of red-shifted emission to the blue-shift in the trailed spectrum. In the middle, a cartoon of the binary system is shown in four orbital phases corresponding to 0.85, 0.95, 1.05, and 1.15. The binary system geometry is to scale.}
\label{fig:ecl}
\end{figure}

\begin{figure*}
\setlength{\unitlength}{1mm}
\resizebox{16cm}{!}{
\begin{picture}(100,53)
%\put (-10,0){\includegraphics[width=12cm, bb= 0 440 1700 1200, clip=]{Figures/DopTom.pdf}}
\put (-10,0){\includegraphics[width=12cm, bb= 0 440 1700 1200, clip=]{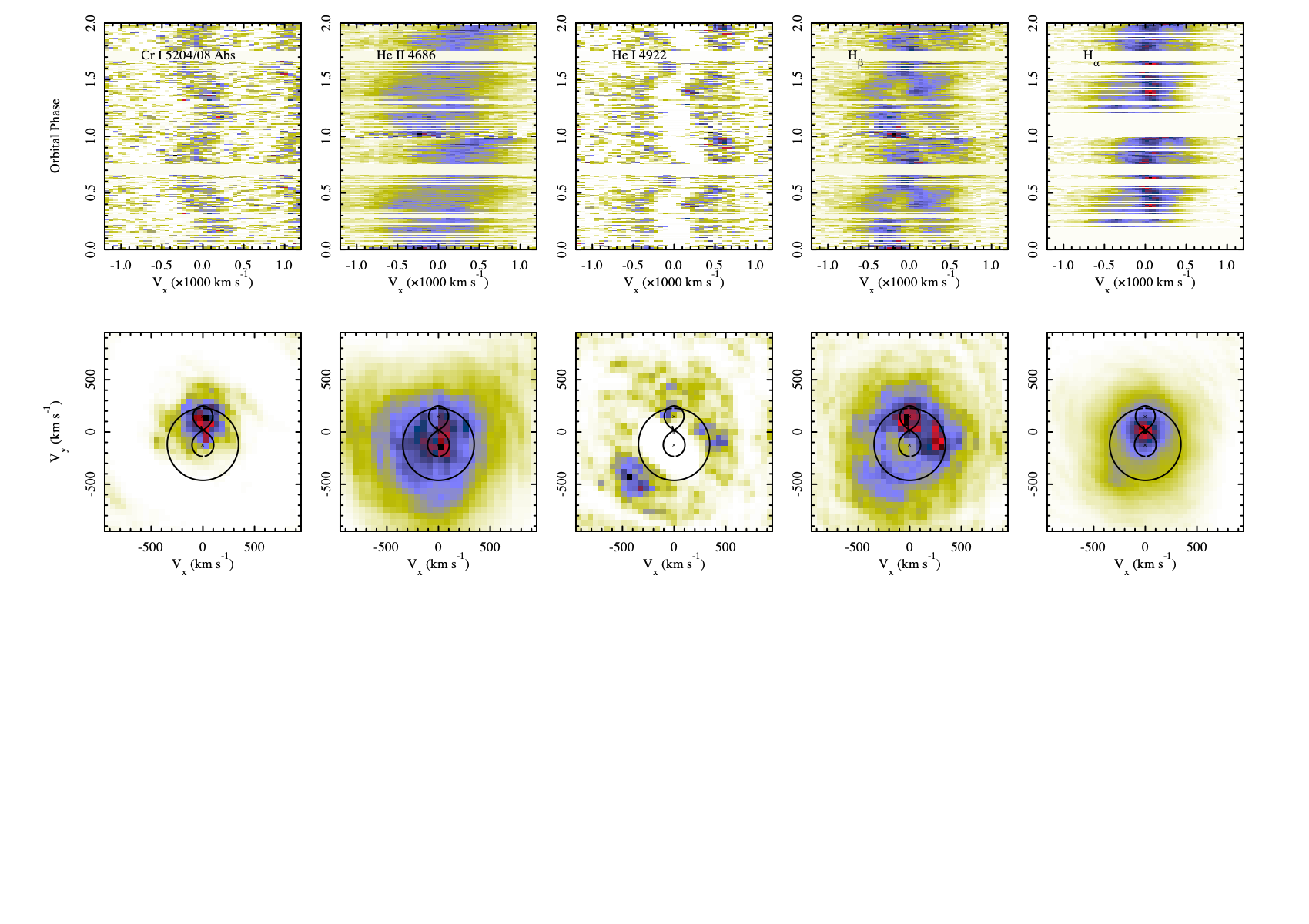}}
\end{picture}}
\caption{The trailed spectra of lines  \ion{Fe}{i}$+$\ion{Cr}{i} in absorption, and emission lines \ion{He}{ii}, \ion{He}{i}, H$\beta$, and H$\alpha$ from left to right are presented in the upper panel. On the vertical axis, two orbital periods are plotted for illustration purposes. In the bottom panel, their corresponding Doppler tomograms are displayed. The description is in the text.}
\label{fig:TrailSpectra}
\end{figure*}

The most direct evidence for disc emission in \obj\ comes from
the eclipse morphology itself. During ingress, the near side of
the rotating disc is progressively occulted, causing a systematic
shift from red-shifted to blue-shifted emission; the reverse
occurs during egress. This signature is clearly seen in the
trailed spectrum of H$\beta$ around the eclipse in
Fig.\,\ref{fig:ecl}, and is independently visible in
\ion{He}{ii}, confirming that the disc contributes substantially
to all major emission lines. The effect becomes more pronounced
as the system brightness decreases during the partial eclipse,
improving the line-to-continuum contrast.

Beyond the eclipse, the disc manifests as two parallel sine
curves flanking the wings of \ion{He}{i}, H$\beta$, and
H$\alpha$ in the trailed spectra (upper panel of
Fig.\,\ref{fig:TrailSpectra}), alongside the trailed spectra
of the \ion{Fe}{i}$+$\ion{Cr}{i} absorption blend and the
\ion{He}{ii}\,$\lambda$4686 emission line shown for comparison.
The corresponding Doppler
tomograms \citep[see][]{Marsh+Horne_1988,Marsh_2005}
are given in the lower panels. The emission-line profiles are
complex and suggest the presence of at least three kinematically
distinct contributing regions, which in principle could be
decomposed into three independent sinusoidal components.
However, given the dominant and spatially extended disc
contribution demonstrated by the eclipse morphology, such a
decomposition would be highly degenerate and is not attempted
here; the Doppler tomograms provide a more robust representation
of the emitting geometry.

The first panels from the left show the Doppler map and trailed
spectrum of the donor absorption line (inverted intensities),
perfectly located within the Roche lobe contour. The following
panels reveal double-peaked disc emission with additional
components producing complex, line-dependent patterns. The
disc emission peak is offset from exact counter-phase with the
donor absorption by $\phi \approx 0.071$, corresponding to
approximately 1.2\,h in this 17\,h system. This is a
substantial and systematic phase shift rather than a minor
asymmetry. It most plausibly arises from the combined effect
of the bright spot at the stream--disc impact point and the
non-axisymmetric disc structure induced by the mass-transfer
stream, both of which are expected to produce a net phase
offset of the integrated disc emission relative to the binary
centre of mass \citep{Marsh+Horne_1988}. The most easily
identifiable additional component is the low-velocity
mass-transfer stream visible in H$\alpha$. A condensation in
the third quadrant of the H$\beta$ tomogram, present at lower
intensity in all lines, is consistent with the stream--disc
impact region. A further condensation in the first quadrant,
partially outside the disc boundary, has no straightforward
identification and may indicate stream overflow or a disc
asymmetry; we refrain from further speculation given the
complexity of the system. The strong \ion{He}{ii} emission
line, originating in the innermost and hottest regions of the
disc, produces a broad, centrally concentrated Doppler map
with large velocity dispersion. Its trailed spectrum shows
the same ingress red-to-blue shift pattern as the Balmer
lines, providing additional confirmation that the inner disc
is eclipsed.

\section{Evolution of \obj}\label{sec:evol}

\begin{table}
\caption{Predicted and observed parameters of \obj.}
\label{Tab:Comparison}
\centering
\setlength\tabcolsep{8.5pt}
\renewcommand{\arraystretch}{1.0}
\begin{tabular}{lcc}
\hline\hline
Parameter & Observed & Predicted  \\
\hline
$P_{\rm orb}$ (hr)    & $17.1094$      & $17.1041$ \\
$R_2$ (\Rsun)         & $1.3-1.4$      & $1.313$ \\
$M_2$ (\Msun)         & $0.57-0.73$     & $0.578$ \\
$T_{\rm eff,2}$ (K)   & $4550-4650$    & $4638$ \\
$\log{}g_2$           & $4$            & $3.964$ \\
\hline
\end{tabular}
\end{table}

\begin{table*}
\centering
\caption{Evolution of a zero-age post-CE binary towards the present-day properties of \obj. The quantities $M_1$ and $M_2$ and Type$_1$ and Type$_2$ are the masses and stellar types of the progenitors of the WD and its companion, respectively. $P_{\rm orb}$ is the orbital period and the last column corresponds to the event occurring to the binary at the time in the first column. The row in which the binary has the present-day properties of \obj~is highlighted in boldface.}
\label{Tab:FormationChannel}
\setlength\tabcolsep{9pt}
\renewcommand{\arraystretch}{1.0}
\begin{tabular}{r c c c c c l}
\hline
\noalign{\smallskip}
 Time  &   $M_{\mathrm{wd}}$    &   $M_2$    & Type$_1$  & Type$_2$ & Orbital Period & Event\\
 (Myr) & (M$_\odot$)&(M$_\odot$) &           &          &  (h)        &      \\
\hline
\noalign{\smallskip}
   0.0000  &  0.8    &  1.100 & WD     & main sequence       &    74.4000  &  initial post-common-envelope binary \\
5430.8294  &  0.8    &  1.098 & WD     & subgiant       &    74.4606  &  secondary becomes a subgiant \\
7114.3189  &  0.8    &  1.096 & WD     & subgiant       &    26.3389  &  onset of RL overflow\\
\textbf{7118.4898}   &  \textbf{0.8}   &  \textbf{0.578} & \textbf{WD}     & \textbf{subgiant}       &    \textbf{17.1041}  &  \textbf{binary looks like \obj} \\
7543.8066  &  0.8    &  0.165 & WD     & proto-WD &    12.9063  &  end of RL overflow \\
\noalign{\smallskip}
\hline
\end{tabular}
\end{table*}

\begin{figure*}
\begin{center}
\includegraphics[width=0.49\linewidth]{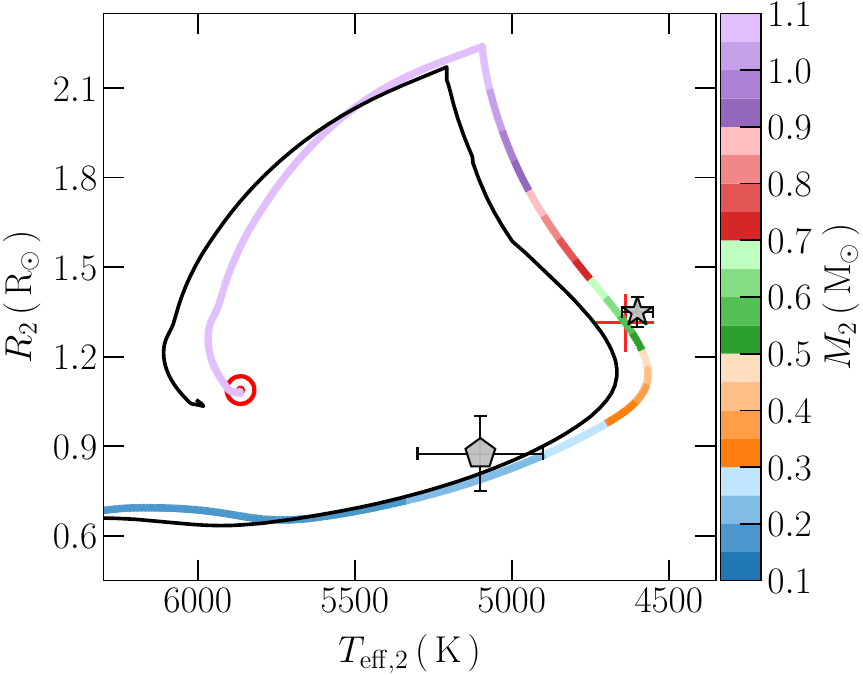}
\includegraphics[width=0.49\linewidth]{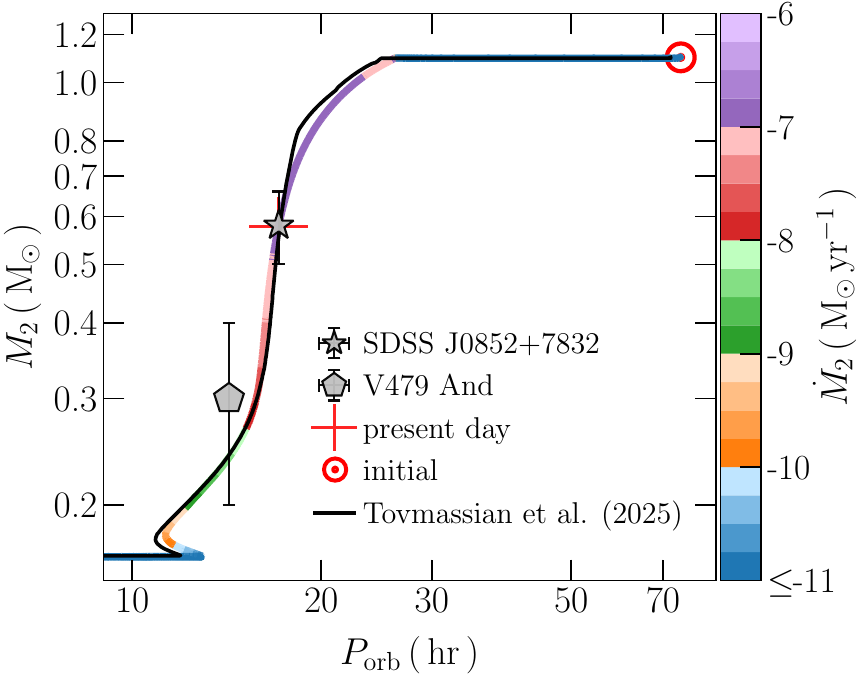}
\end{center}
\caption{Post-CE evolution of the radius with the effective temperature of the donor in \obj, colour-coded by its mass (left panel) and of the mass of the donor with the orbital period, colour-coded by the mass transfer rate (right panel). The red solar symbol indicates the properties at the moment the WD was formed (i.e., just after CE evolution), and the thick red cross indicates the properties at the moment the orbital period is the same as observed. The grey markers correspond to the properties derived from observations for \obj~(star, this work) and \VAnd~(pentagon, \citealt{Tovmassian_2025}).}
\label{Fig:Model}
\end{figure*}

We can use the parameters of the donor in \obj~to search for reasonable evolutionary sequences that can explain its properties, following the steps we have already taken for \VAnd~and \VSgr~\citep{Tovmassian_2025}.
We used the MESA code \citep[][r15140]{Paxton_2011,Paxton_2013,Paxton_2015,Paxton_2018,Paxton_2019,Jermyn_2023} to calculate binary evolution of the \obj. As a result, we successfully found a reasonable model for \obj, which is illustrated in Fig.\,\ref{Fig:Model}.
The properties we predict at the observed orbital period are in very reasonable agreement with the observed (Table~\ref{Tab:Comparison}), and our evolutionary sequence is summarized in Table~\ref{Tab:FormationChannel}.
The model shows that Roche-lobe overflow begins at $P_{\rm orb}\approx26.3$\,h and the present-day configuration is reached $\sim4$\,Myr later, during which the donor loses $\sim0.52$\,\Msun. This corresponds to a mean mass-transfer rate of 
$\sim\!1.2\times10^{-7}$\,\Msunyr\ over this phase, declining 
to the current rate $\dot{M} = 7.1(8)\times10^{-8}$\,\Msunyr\ 
deduced from light curve modelling (Table~\ref{tab:BestPar}), 
as the system settles towards a quasi-stable mass-transfer regime.
A summary of the main assumptions is provided in what follows, but more details are found in \citet{Belloni+Schreiber_2023}\,\footnote{\href{https://zenodo.org/records/8279474}{https://zenodo.org/records/8279474}} and \citet{Tovmassian_2025}.

The MESA equation of state is a blend of the OPAL \citep{Rogers_2002}, SCVH \citep{Saumon_1995}, FreeEOS \citep{Irwin_2004}, HELM \citep{Timmes_2000}, PC \citep{Potekhin_2010}, and Skye \citep{Jermyn_2021} equations of state.
Nuclear reaction rates are a combination of rates from NACRE \citep{Angulo_1999}, JINA REACLIB \citep{Cyburt_2010}, plus additional tabulated weak reaction rates \citep{Fuller_1985,Oda_1994,Langanke_2000}.
Screening is included via the prescription of \citet{Chugunov_2007} and thermal neutrino loss rates are from \citet{Itoh_1996}.
Electron conduction opacities are from \citet{Cassisi_2007} and radiative opacities are primarily from OPAL \citep{Iglesias_1993,Iglesias_1996}, with a high-temperature Compton-scattering dominated regime calculated using the equations of \citet{Buchler_1976}.
We assumed a metallicity of ${Z=0.015}$, which is required to explain the low effective temperature and large radius of the donor star.
We treated convective regions using the scheme by \citet{Henyey_1965} for the mixing-length theory, adopting a mixing length of $1.5\,H_{\rm p}$, where $H_{\rm p}$ is the pressure scale height at the convective boundary.
We assumed the grey Eddington T($\tau$) relation to calculate the outer boundary conditions of the atmosphere \citep[][their sect.~5.3]{Paxton_2011}.
We adopted a uniform opacity that is iterated to be consistent with the final surface temperature and pressure at the base of the atmosphere.
For low-temperature radiative opacities, we adopted the tables from \citet{Ferguson_2005}.
We also used the nuclear network \texttt{cno$\_$extras.net}, which accounts for the nuclear reactions of the carbon-nitrogen-oxygen-hydrogen burning cycle.

The main difference from our previous works is that we accounted for the impact of helium novae \citep{1989ApJ...340..509K} on WD mass growth.
For hydrogen accretion, we included the critical accretion rates calculated by \citet[][]{Wolf_2013}, as before.
For helium accretion, i.e., the helium produced by stable hydrogen burning, we estimated the amount, which depends on the rate at which hydrogen is converted to helium, adopting the models from \citet{Kato_2004} and \citet{Wu_2017}.
An essential ingredient in the simulations carried out by \citet{Belloni+Schreiber_2023} and \citet{Tovmassian_2025}, which is also incorporated here, is the use of the CARB model \citep{Van+Ivanova_2019} to account for strong angular momentum loss via magnetic braking.

\section{Discussion}

The long-period eclipsing binary SDSS\,J085210.48+783246.6 adds
a new member to the small but growing family of cataclysmic
variable-like systems with significantly evolved donor stars.
The derived stellar and binary parameters --- in particular, the
$17.1$\,h orbital period, the K3\,IV-type Roche-lobe-filling
donor with $M_2 = 0.6\pm0.1$\,\Msun\ and $R_2 \simeq
1.4$\,\Rsun, and the presence of an accretion disc
(with a disc luminosity $L_{\rm disc}\sim G M_{\rm WD}\dot{M}/R_{\rm WD}\sim4\times10^{32}$\,erg\,s$^{-1}$, comparable to the donor's bolometric luminosity of $\sim3\times10^{32}$\,erg\,s$^{-1}$ for $T_{\rm eff}=4500$\,K and $R_2\simeq1.4$\,\Rsun) around
a $\sim\!0.8$\,\Msun\ white dwarf --- demonstrate that
\obj\ is not a typical CV. Its donor star possesses a luminosity
and temperature characteristic of a subgiant, yet its mass is
markedly lower than that of an isolated star of similar spectral type.
This is a direct consequence of a previous phase of
thermal-timescale mass transfer that stripped much of the donor
envelope, after which the system settled into a quasi-stable
mass-transfer regime characteristic of an advanced evolutionary
stage.

The donor parameters inferred for \obj\ are supported by several
independent approaches. The radius derived from the spectral energy distribution and Gaia distance is consistent with that
required for a Roche-lobe-filling donor at the observed orbital
period. The $R_2$--$M_2$ diagnostic diagram provides a graphical
representation of this constraint by comparing the observed donor
radius with the Roche-lobe size expected for a binary with the
measured orbital period and different assumed white dwarf masses
\citep{Tovmassian_2025}. The intersection between these relations
yields a donor mass of $\sim0.68^{+0.05}_{-0.11}\,\Msun$, in good agreement with
the values obtained from light-curve modelling and from the
evolutionary models.

It is noteworthy that the donor parameters derived from four
independent methods --- SED fitting, the $R_2$--$M_2$ diagnostic
diagram, light curve modelling, and the \textsc{MESA} evolutionary
models --- are mutually consistent within their respective
uncertainties (Tables\,\ref{tab:BestPar} and \ref{Tab:Comparison}). Despite relying on different assumptions and
levels of model dependence, all approaches converge on a donor
mass of $\sim0.6$--$0.7\,\Msun$ and a radius of
$\sim1.4\,\Rsun$. This consistency lends confidence to the
overall characterization of the system despite the discrete
nature of the model grids employed.

Our detailed modelling with \textsc{MESA} shows that
\obj\ follows an evolutionary track similar to that of the
long-period system V479\,And \citep{Tovmassian_2025}, differing
mainly in being at an earlier point along that sequence. Both
binaries are reproduced by post-common-envelope progenitors with
comparable component masses under strong magnetic braking
consistent with the CARB prescription of
\citet{Van+Ivanova_2019}. Within this framework, \obj\ will
naturally evolve towards a V479\,And-like configuration as mass
transfer continues and the orbital period shortens.

The two systems differ, however, in ways that underscore the
diversity of outcomes within this evolutionary class. V479\,And
shows clear chemical abundance anomalies --- including
CNO-processed material at the surface --- and observational
signatures consistent with a synchronously rotating, strongly
magnetized white dwarf \citep{Tovmassian_2025}. The CNO surface
anomalies in V479\,And indicate that the donor's convective
envelope has dredged up material from layers previously processed
by hydrogen burning, consistent with a more advanced stage of
nuclear evolution than that inferred for \obj. For \obj, by contrast, we find no evidence for a strongly magnetized white
dwarf: the emission-line profiles and Doppler tomograms are
consistent with disc-dominated accretion in a non-magnetic
nova-like system with a hot, flared disc \citep{Knigge_2000}.
These differences highlight the importance of detailed
spectroscopic and magnetic diagnostics in characterizing
long-period CVs with evolved donors.

What the present system does not yet tell us is where this sequence ultimately leads. The \textsc{MESA} models predict
that \obj\ will continue losing mass and contract towards
shorter orbital periods, eventually detaching once the hydrogen envelope is exhausted. The timescale on which the resulting binary
subsequently evolves into a helium-transferring AM\,CVn system determines whether the system can be considered as representing progenitors of currently observed AM\,CVn systems.
If the orbital period at detachment is short
enough for gravitational-wave emission to drive the components
back into contact within a Hubble time the latter would be the case.

However, our evolutionary track predicts detachment at
$P_{\rm orb}\approx12.9$\,h (Table\,\ref{Tab:FormationChannel}), which means that \obj\ is different from the progenitors of currently observed AM\,CVn systems, as the time to restart mass transfer exceeds a Hubble time. For the predicted component masses at detachment ($M_{\rm WD}\approx0.8$\,\Msun, $M_2\approx0.165$\,\Msun), the gravitational-wave merger timescale from $P_{\rm orb}\approx12.9$\,h is $\tau_{\rm GW}\sim G^3M_1M_2(M_1+M_2)/(48c^5a^4)\sim10^{12}$\,yr \citep{Belloni+Schreiber_2023}, far exceeding the Hubble time. We therefore conclude that, while the predicted evolution of \obj\ resembles that predicted by \citet{Belloni+Schreiber_2023} for the formation of AM\,CVn, with respect to currently observed populations,
\obj\ can be considered a progenitor of close double degenerate binaries.

What is established here is more immediate: the existence of a
population of long-period, mass-transferring binaries with
undermassive, nuclear-evolved donors that cannot be accommodated
within the standard CV framework, and whose properties are
naturally explained by strong magnetic braking acting on
slightly evolved donors. Each new member of this class ---
\obj, V479\,And, V1082\,Sgr --- adds observational weight to the hypothesis that this channel is a significant, previously
underestimated pathway in compact binary evolution.

\section{Conclusions}

SDSS\,J085210.48+783246.6 represents another example of a
long-period interacting binary with a significantly evolved
donor star. The system has an orbital period of $17.1$\,h and
contains a K3\,IV Roche-lobe-filling donor with
$M_2 \sim 0.6$--$0.7\,\Msun$ and $R_2 \simeq 1.4\,\Rsun$
transferring mass onto a $\sim0.8\,\Msun$ white dwarf.

The donor is markedly undermassive for its spectral type,
indicating that it has undergone substantial envelope stripping
during a previous phase of thermal-timescale mass transfer. The
consistency between several independent parameter determinations
--- SED fitting, the $R_2$--$M_2$ diagnostic diagram, light-curve modelling, and evolutionary modelling --- provides a coherent picture of the system.

Evolutionary calculations suggest that \obj\ is presently at an earlier stage of the evolutionary sequence that includes systems such as V479\,And. Together with V1082\,Sgr and related objects, it supports the existence of a class of long-period CV-like binaries with evolved donors whose formation requires strong magnetic braking and significant prior nuclear evolution of the secondary star.

\section*{Acknowledgements}

GT was supported by grants  IN109723 from the Programa de Apoyo a Proyectos de Investigación e Innovación Tecnológica (DGAPA-PAPIIT).
DB acknowledges support from the S\~{a}o Paulo Research Foundation (FAPESP), Brazil, Process Numbers {\#2024/03736-2} and {\#2025/00817-4}.
SZh acknowledges grant  IN105826, and JE was supported by grant IN113723, both from DGAPA-PAPIIT.
This work is supported by the National Natural Science Foundation of China (NSFC, Nos. 12288102, 12125303), the Strategic Priority Research Program of the Chinese Academy of Sciences (grant Nos. XDB1160303, XDB1160300, XDB1160000, XDB1160200, XDB1160201), the CAS Project for Young Scientists in Basic Research (YSBR-148), the Yunnan Revitalization Talent Support Program ``YunLing Scholar'' project, International Centre of Supernovae (ICESUN), Yunnan Key Laboratory of Supernova Research (No. 202505AV340004),  the New Cornerstone Science Foundation through the XPLORER PRlZE, and the Yunnan Revitalization Talent Support Program-Science \& Technology Champion Project (No. 202305AB350003).
We are grateful to Dr.\ U. Munari, who facilitated the observations at the Osservatorio Astrofisico di Asiago.

This paper includes data collected with the TESS mission, obtained from the MAST data archive at the Space Telescope Science Institute (STScI). Funding for the TESS mission is provided by the NASA Explorer Program. STScI is operated by the Association of Universities for Research in Astronomy, Inc., under NASA contract NAS 5–26555.
Funding for the Sloan Digital Sky Survey V has been provided by the Alfred P. Sloan Foundation, the Heising-Simons Foundation, the National Science Foundation, and the Participating Institutions. SDSS acknowledges support and resources from the Center for High-Performance Computing at the University of Utah. SDSS telescopes are located at Apache Point Observatory, funded by the Astrophysical Research Consortium and operated by New Mexico State University, and at Las Campanas Observatory, operated by the Carnegie Institution for Science. The SDSS website is \url{www.sdss.org}.
SDSS is managed by the Astrophysical Research Consortium for the Participating Institutions of the SDSS Collaboration, including the Carnegie Institution for Science, Chilean National Time Allocation Committee (CNTAC) ratified researchers, Caltech, the Gotham Participation Group, Harvard University, Heidelberg University, The Flatiron Institute, The Johns Hopkins University, L'Ecole polytechnique f\'{e}d\'{e}rale de Lausanne (EPFL), Leibniz-Institut f\"{u}r Astrophysik Potsdam (AIP), Max-Planck-Institut f\"{u}r Astronomie (MPIA Heidelberg), Max-Planck-Institut f\"{u}r Extraterrestrische Physik (MPE), Nanjing University, National Astronomical Observatories of China (NAOC), New Mexico State University, The Ohio State University, Pennsylvania State University, Smithsonian Astrophysical Observatory, Space Telescope Science Institute (STScI), the Stellar Astrophysics Participation Group, Universidad Nacional Aut\'{o}noma de M\'{e}xico, University of Arizona, University of Colorado Boulder, University of Illinois at Urbana-Champaign, University of Toronto, University of Utah, University of Virginia, Yale University, and Yunnan University.

Some/all of the data presented in this paper were obtained from the Multimission Archive at the Space Telescope Science Institute (MAST). STScI is operated by the Association of Universities for Research in Astronomy, Inc., under NASA contract NAS5-26555. Support for MAST for non-HST data is provided by the NASA Office of Space Science via grant NAG5-7584 and by other grants and contracts.

\vspace{3mm}
\noindent\textit{Software and facilities:}
This work made use of
\textsc{Astropy} \citep{AstropyCollaboration_2013,
AstropyCollaboration_2018},
\textsc{MESA} \citep[][r15140]{Paxton_2011, Paxton_2013,
Paxton_2015, Paxton_2018, Paxton_2019, Jermyn_2023},
\textsc{pamela} and \textsc{molly} \citep{Marsh1989},
and \textsc{iraf}.
The following facilities were used:
\textit{Gaia}, SDSS\,V, \textit{Swift} (XRT and UVOT),
TESS, ZTF, ASAS-SN, ATLAS,
ING\,2.5\,m (INT/IDS), and Asiago\,1.22\,m.

\section*{Data Availability}

The data analysed in this work can be made available upon reasonable request to the authors.

\bibliographystyle{mnras}
\bibliography{sdss0852}

\bsp
\label{lastpage}
\end{document}